\documentclass[10pt,journal,twoside]{IEEEtran}
\usepackage{color}
\usepackage{hyperref} 
\usepackage{graphicx, graphics, theorem, times, amsfonts, amsmath, amssymb}
\usepackage{subfig}
\usepackage{multirow}

\input{mysymbol.sty}

\usepackage{cite}

\ifCLASSINFOpdf
\else
\fi
\usepackage{amsmath}
\usepackage{url}

\begin{document}
%
\title{Learning to Model the Relationship Between Brain Structural and Functional Connectomes}
%
%
%

\author{Yang~Li,~\IEEEmembership{Student Member,~IEEE,}
        Gonzalo~Mateos,~\IEEEmembership{Senior Member,~IEEE,}
        and~Zhengwu~Zhang,~\IEEEmembership{Member,~IEEE}
\thanks{This work was supported in part by NSF Awards under
Grants CCF-1750428, CCF-1934962, and ECCS-1809356. \emph{(Corresponding author: Gonzalo Mateos.)}}
\thanks{Y. Li and G. Mateos are with the Dept.
of Electrical and Computer Engineering, University of Rochester, Rochester,
NY, 14627 USA. E-mails: \{yli131,gmateosb\}@ur.rochester.edu.}
\thanks{Z. Zhang is with the Dept.
of Statistics and Operations Research, University of North Carolina at Chapel Hill. E-mail: zhengwu$\_$zhang@unc.edu.}
\thanks{Part of the results in this paper were presented at the \textit{2020 ICASSP Conference}~\cite{li2020supervised}.}}

\maketitle

\begin{abstract}
Recent advances in neuroimaging along with algorithmic innovations in statistical learning from network data offer a unique pathway to integrate brain structure and function, and thus facilitate revealing some of the brain's organizing principles at the system level. In this direction, we develop a supervised graph representation learning framework to model the relationship between brain structural connectivity (SC) and functional connectivity (FC) via a graph encoder-decoder system, where the SC is used as input to predict empirical FC. A trainable graph convolutional encoder captures direct and indirect interactions between brain regions-of-interest that mimic actual neural communications, as well as to integrate information from both the structural network topology and nodal (i.e., region-specific) attributes. The encoder learns node-level SC embeddings which are combined to generate (whole brain) graph-level representations for reconstructing empirical FC networks. The proposed end-to-end model utilizes a multi-objective loss function to jointly reconstruct FC networks and learn discriminative graph representations of the SC-to-FC mapping for downstream subject (i.e., graph-level) classification. Comprehensive experiments demonstrate that the learnt representations of said relationship capture valuable information from the intrinsic properties of the subject's brain networks and lead to improved accuracy in classifying a large population of heavy drinkers and non-drinkers from the Human Connectome Project. Our work offers new insights on the relationship between brain networks that support the promising prospect of using graph representation learning to discover more about human brain activity and function. 
\end{abstract}


%
\IEEEpeerreviewmaketitle


\section{Introduction}
%
%
%

\IEEEPARstart{T}{he} human brain is a complex yet efficient information processing network, whose distributed organization entails different regions conducting individual tasks while actively interacting with each other~\cite{van2010exploring}. 
This \emph{integrative} nature of brain function along with recent advances in neuroimaging, motivate well the adoption of graph-centric signal and information processing tools to study the interplay between brain structure (of neural connections often referred to as the structural connectome)~\cite{fornito2016fundamentals} and functional activity~\cite{power2011functional}. Indeed, understanding the underpinnings of brain function is one of the fundamental scientific challenges of this century.

Brain connectivity broadly consists of two types of networks. \emph{Structural connectivity} (SC) has to do with anatomical tracts of axonal bundles~\cite{bullmore2009complex}, which can be extracted using tractography algorithms applied to diffusion magnetic resonance imaging (dMRI). On the other hand, \emph{functional connectivity} (FC) represents pairwise statistical correlation between activation signals in various brain regions of interest (RoIs)~\cite{power2011functional, bullmore2009complex}. These so-termed blood oxygen level-dependent (BOLD) signals are measured by functional MRI (fMRI). The interplay between SC and FC is of great importance and a timely area of research in network neuroscience. There is consensus that further understanding of such coupling could offer new and important insights on the inner working mechanics of the brain~\cite{sporns2010networks,abdelnour2014network, li2019identifying}. Previous studies have revealed that FC correlates with SC at an aggregate level\cite{honey2009predicting}. Although FC is shaped by the actual anatomical connections within the brain, strong functional connections exist between brain regions that are not directly linked by structural connections, or have rather limited anatomical pathways\cite{goni2014resting}. These findings provide strong evidence that functional interactions between brain regions depend on both direct and indirect anatomical connections\cite{stam2016relation}, and motivate multiple previous works in predicting FC from SC~\cite{honey2010can, honey2009predicting, goni2014resting,GlobalSIP19_YangLiEncoderDecoder}, and in estimating SC from FC~\cite{li2019identifying}. Recently, \cite{sarwar2021structure} developed a deep learning model to predict FC from SC, and reported that the structure-function coupling in human brain networks is substantially tighter than previously suggested. 

In this paper, we model the structure-function relationship in the brain by learning low-dimensional representations of the SC-to-FC mapping using graph convolutional networks (GCNs)\cite{kipf2016semi,duvenaud2015convolutional,gama2018convolutional}. This way, we aim at better exploiting the relational inductive biases present in brain connectome data. 
While the benefits of graph representation learning (GRL) methods have been well documented in applications ranging from recommender systems\cite{ying2018graph} to social network analysis\cite{hamilton2017inductive}, their impact to network neuroscience is yet to fully materialize~\cite{yue2020graph, ktena2018metric}. In the context of brain network analysis, it is not apparent how to generalize inductive GRL models with learnable parameters shared among a large cohort of subjects; and we explore this avenue here. Specifically, we propose and test a supervised GRL framework to jointly learn: (i) low-dimensional \emph{node} embeddings generated from SC networks to reconstruct empirical FC; and (ii) learn \emph{graph} embeddings to represent the whole graphs for subject-level classification. 

Our overall architecture can be viewed as a graph encoder-decoder system inspired by the graph autoencoder model\cite{kipf2016variational}; see also~\cite{chami2020machine}. For each RoI in the input SC graph, the encoder outputs low-dimensional node embeddings that integrate both nodal attributes (when available) and the local graph topology information. To aggregate information from multiple hops within the network, layered graph convolution operations interleaved with point-wise nonlinearities are used to compute nodal features. 
This makes the GCN a proper encoder model to generate connectome embeddings that capture indirect connections inherent to brain networks, and the SC-FC relationship in particular. Whole graph embeddings are then obtained by applying certain global pooling mechanisms on the node embeddings\cite{cheung2019pooling}.  
All in all, we propose a multi-task learning model with the dual goal of reconstructing brain FC from input SC data, and performing subject classification using the learnt graph embeddings as features. The graph estimation/regression task learns a parsimonious representation of the population-level SC-FC relationship, and the classification task uses subject labels as additional inputs for supervised learning. Accordingly, the model strikes the right balance between deciphering population patterns that shape the SC-FC coupling, and distilling subject-level variability to facilitate graph classification with regard to drinking habits

We train and test our graph encoder-decoder model on a neuroimaging dataset of $412$ subjects from the Human Connectome Project (HCP)~\cite{HCP}, and obtain satisfactory FC reconstruction performance as well as subject classification accuracy (classes are $191$ non-drinkers and $221$ heavy drinkers). The proposed GRL pipeline outperforms baseline auto-encoder methods that only rely on SC or FC, supporting the thesis that SC-to-FC mapping provides valuable information to better discriminate among the classes beyond static graph topology, specifically in the study of alcohol's effect on human brain. Via statistical tests on the reconstructed FC networks, brain sub-graphs exhibiting significant difference between groups are also unveiled. All in all, the novelty of our work is in learning low-dimensional, latent representations that capture the intrinsic attributes of the brain SC-FC coupling across a large population, while being discriminative for subject-level classification. This paper is not about fundamental innovations in the GRL space, but instead it offers a novel application of this framework to neuroimaging data analysis with the distinct perspective of modeling the SC-FC relationship in the brain.


\subsection{Related works}\label{ssec:related_work}


GCNs are versatile signal and information processing architectures\cite{kipf2016semi,duvenaud2015convolutional,gama2018convolutional}, which comprise stacked layers of graph (convolutional) filters followed by point-wise nonlinearities; see\cite{chami2020machine,gama2020spmag,bronstein2017geometric, errica2020fair,wu2021comprehensive, gama2018convolutional} for recent surveys and the references therein. From early spectral convolutions\cite{bruna2013spectral, kipf2016semi} to distributed implementations of (equivalent) shift-invariant polynomial graph filters\cite{gama2020spmag,hamilton2017inductive, xu2018powerful}, GCNs 
integrate information from both the graph topology and nodal attributes to learn representations of network data. Indeed, the GRL paradigm is to learn low-dimensional embeddings of individual vertices, edges, or the graph itself~\cite{hamilton2020book,chami2020machine,hamilton2017representation,gutierrez2019unsupervised}, which can then be used in e.g., (semi-supervised) node classification~\cite{kipf2016semi}, link prediction~\cite{zhang2018linkpred}, graph clustering\cite{tsitsulin2020graph,tian2014learning}, and graph classification\cite{narayanan2017graph2vec}. Recently, GRL ideas have permeated to neuroimaging data analysis for behavioral state classification~\cite{wang2021generalizable}, to study the relationship between SC and FC\cite{rosenthal2018mapping,GlobalSIP19_YangLiEncoderDecoder}, and to extract representations for subject classification\cite{kim2020understanding,zhang2018multi,parisot2018disease}. Although these prior works have probed the area of brain connectomics in several novel directions and achieved solid performance in multiple regression or classification tasks, they mostly rely on a single type of brain network and use feed-forward models to predict subject labels from input graphs. Our different approach is to distill actionable information from the relationship between FC and SC. Instead of learning representations of a certain type of brain network, we propose a multi-task GRL pipeline to model the SC-FC coupling, thus seamlessly integrating brain structure and function in a principled way.


\subsection{Summary of contributions}\label{ssec:contributions}
In summary, we develop a GCN-based supervised encoder-decoder system to learn parsimonious latent-space representations of the SC-FC relationship from HCP neuroimaging data. 
Different from the surveyed deep learning approaches for brain network analyses, to the best of our knowledge this is the first work to explore learnt representations of the mapping between two brain connectivity modalities -- among which a tight coupling is well documented~\cite{honey2009predicting, honey2010can, rosenthal2018mapping}. Our model can extract valuable and interpretable information from the intrinsic properties of the subjects' brain networks, leading to an improved performance trade-off between FC reconstruction error and graph classification accuracy relative to baseline methods. We also identified brain sub-networks exhibiting significant functional connectivity differences among heavy drinkers and non-drinkers. Such data-driven findings may serve as explanations behind the improved classification performance. They could also offer additional insights to guide follow-up studies aiming at understanding of alcohol's effects to the brain.

Unlike the conference precursor~\cite{li2020supervised} to this paper, here we thoroughly examine different GCN-based encoder architectures with multiple layers and various number of filters per layer. Inspired by\cite{xu2018powerful}, we also find that concatenating the outputs from each GCN layer as a readout function improves model performance.  As a final component in our architectural design, we explore various global (mean/max/sum) pooling mechanisms to form graph-level representations from the collection of node embeddings. Accordingly, the representation space of the resulting model is markedly richer than the shallow GCN encoder in~\cite{li2020supervised}. Preliminary tests in~\cite{li2020supervised} did not include comparisons with FC or SC auto-encoder baselines, results which here allow us to justify our claim on the merit of learning SC-FC representations (that inherently integrate brain structure and function).  In a broader context, the proposed multi-task GRL framework could be applied to other domains beyond network neuroscience. We believe it is especially well suited to combine dynamic or multi-view (multi-aspect) network data, while being capable to extract interpretable and discriminative patterns for supervised learning applications.  


\section{Graph neural networks background}\label{sec:gnn_background}
In this section, the required background of graph neural networks (GNNs) is briefly outlined. We first present the encompassing message passing formalism used for a broad class of GNNs~\cite{xu2018powerful}. As the GCN\cite{kipf2016semi} is adopted for the encoder in the GRL pipeline, a brief review of graph convolutional models is provided as well; see also\cite{chami2020machine, bronstein2017geometric, errica2020fair,wu2021comprehensive, gama2018convolutional} for comprehensive surveys with further details.


\subsection{Message passing on graphs}\label{ssec:message_passing}

Consider a weighted, undirected graph denoted by $G := (\ccalV, \ccalE)$, where $\ccalV=\{1,\ldots,N\}$ is a set of $N$ vertices corresponding to brain RoIs and $\ccalE\subseteq \ccalV\times \ccalV$
are edges in which $(i,j)\in\ccalE$ is a structural connection
joining $i$ and $j$. We define the neighborhood of vertex $i$ as
$\ccalN(i)=\{j\in \ccalV :  (i,j)\in \ccalE\}$. The symmetric adjacency matrix $\bbA\in\mbR_+^{N\times N}$ has entries $A_{ij}\geq 0$ representing the structural connection strengths between RoIs $i$ and $j$; $A_{ij}=0$ indicates $(i,j)\notin \ccalE$. Henceforth we will use $\bbSigma$ to denote the $N\times N$ adjacency matrix of the FC network, making explicit this is a correlation matrix (and different from $\bbA$). Network data models often augment $G$ with a vertex-valued signal $\bbx\in\reals^N$, where $x_i$ denotes the signal value at node $i$, for example, the nodal attributes (or features) on the brain FC (or SC) network. Extensions to feature vectors $\bbx_i\in\reals^d$ per node are straightforward, leading to a signal matrix $\bbX\in\reals^{N\times d}$ that serves as input to the GNN.

For a broad class of GNN models~\cite{kipf2016semi,gama2018convolutional,xu2018powerful}, the $\ell$-th layer computations to update node representations (or embeddings) $\bbx_i^{(\ell)}$, $i\in\ccalV$, are often implemented via local message passing operations to aggregate local neighborhood information in $G$; see e.g.,~\cite{gilmer2017neural, garg2020generalization}. These encompassing message-passing neural networks (MPNNs) can be mathematically described in terms of two simple steps per layer. The first one is a one-hop neighborhood aggregation step, namely
%
%
\begin{equation}\label{E:agg}
	\bbalpha_i^{(\ell)} = \text{AGGREGATE}^{(\ell)}\left(\{\bbx_j^{(\ell-1)}: j \in \ccalN(i)\}\right),
\end{equation}
where we initialize $\bbx_i^{(0)}=\bbx_i$ for all $i\in\ccalV$. The aggregated neighborhood information $\bbalpha_i^{(\ell)}$ is then combined with node $i$'s own representation $\bbx_i^{(\ell-1)}$, to yield the updated embedding
\begin{equation}\label{E:combine}
	\bbx_i^{(\ell)} = \text{COMBINE}^{(\ell)}\left(\bbx_i^{(\ell-1)},\bbalpha_i^{(\ell)}\right).
\end{equation}
In deep models that stack $L$ such layers, $\bbx_i^{(L)}$ embeds information within node $i$'s $L$-hop neighborhood. The $\text{AGGREGATE}$ and $\text{COMBINE}$ functions are central to MPNN architectures; in Section \ref{ssec:gcn} we detail the choices of these functions for the subsumed GCNs. $\text{AGGREGATE}$ is typically carried out via multiplication with a connectivity matrix $\bbS\in \reals^{N\times N}$ that encodes $G$'s topology; also known as graph-shift operator (GSO) in the graph signal processing (GSP) parlance~\cite{ortega2018graph}. The $\text{COMBINE}$ function often includes a point-wise nonlinear activation function such as ReLU, in addition to trainable parameters in filters we also review in Section \ref{ssec:gcn}.

When needed, a representation $\bbx_G$ for the whole graph $G$ is obtained via global pooling of the node embeddings in the last layer, that is
\begin{equation}\label{E:gpool}
	\bbx_G = \text{READOUT}\left(\{\bbx_i^{(L)}: i \in \ccalV\}\right).
\end{equation}
The $\text{READOUT}$ function can be a simple function such as summing or averaging, or parameterized pooling methods\cite{hamilton2017inductive,xu2018powerful}. Node-level representations $\{\bbx_i^{(L)}\}_{i\in\ccalV}$ can be used for node classification or clustering, while the graph-level representations in \eqref{E:gpool} are used for graph classification; e.g.,~\cite{hamilton2020book}.


\subsection{Graph convolutional networks}\label{ssec:gcn}

The main building block of spectral graph theory is the graph Laplacian matrix defined as $\bbL\!:= \bbD\!-\!\bbA$, where $\bbD:=\textrm{diag}(\bbA\mathbf{1})$ is the diagonal matrix of nodal degrees. The Laplacian $\bbL$ is symmetric and positive semidefinite. Accordingly, it can be decomposed as $\bbL = \bbU \bbLambda \bbU^\top$, where $\bbU\in\mbR^{N\times N}$ denotes the set of orthonormal eigenvectors and the diagonal matrix $\bbLambda$ contains all the non-negative eigenvalues $0=\lambda_1\leq \lambda_2\leq \ldots\leq \lambda_N:=\lambda_{\max}$. The eigenvectors in $\bbU$ serve as Fourier modes to  decompose graph signals~\cite{ortega2018graph}. Specifically, the graph Fourier transform (GFT) of $\bbx$ is defined as $\tbx := \bbU^\top\bbx$, where $\tbx=[\tilde{x}_1,\ldots,\tilde{x}_N]^\top$ collects the graph spectral coefficients of $\bbx$ at frequencies given by the Laplacian eigenvalues~\cite{ortega2018graph,huang2018graph}. 

To process signals $\bbx$ efficiently while incorporating $G$'s topology information, one defines the graph convolution as
\begin{equation}\label{E:graphconv}
	\bbH\bbx = \sum_{k=0}^{K}h_k\bbL^k\bbx,
\end{equation}
where $\bbH := \sum_{k=0}^{K}h_k\mathbf{L}^k$ is a graph filter with coefficients $\bbh := [h_0,\dotsc,h_K]^\top$. The graph convolution operation aggregates signal values from $K$-hop neighborhoods in $G$. Indeed, computation of $\bbL^K\bbx=\bbL(\bbL^{K-1}\bbx)$ entails a sequence of $K$ one-hop aggregations (shifts or diffusions) through multiplication with $\bbL$. Graph filters can be more generally defined as polynomials of GSOs  beyond the combinatorial Laplacian $\bbL$, for instance the adjacency matrix $\bbA$ or suitable normalized variants of the aforementioned algebraic graph topology descriptors~\cite{ortega2018graph,gama2020spmag}. With this interpretation, the analogy of \eqref{E:graphconv} with temporal convolutions implemented via (shift and sum) finite impulse response (FIR) filters should be apparent. Graph convolutions are thus linear shift-invariant operators because $\bbH \bbS = \bbS\bbH$, where $\bbS=\bbL$ in \eqref{E:graphconv}. The naturalness of the definition can be further appreciated in the graph frequency domain, since the GFT of the filtered signal \eqref{E:graphconv} becomes
\begin{equation}\label{E:graphconv_spectral}
	\bbU^\top\bbH\bbx = \bbU^\top\sum_{k=0}^{K}h_k\left(\bbU \bbLambda \bbU^\top\right)^k\bbx = \left(\sum_{k=0}^{K}h_k\bbLambda^k\right)\tbx.
\end{equation}
Notice how the GFT diagonalizes the filter, so that the convolution becomes the element-wise (per frequency $\lambda_i$) multiplication between the filter's frequency response $\tilde{h}_i := \sum_{k=0}^{K}h_k\lambda_i^k$ and the signal's GFT coefficient $\tilde{x}_i$.

GCNs broadly consist of stacked layers of learnable graph convolutional filters and point-wise nonlinearities. Convolutional NNs for network data can be traced to~\cite{bruna2013spectral}, where the filter's frequency response $\tbh^{(\ell)}=[\tilde{h}_1^{(\ell)},\ldots,\tilde{h}_N^{(\ell)}]^\top$ at each layer $\ell$ is learnt using stochastic gradient descent. While intuitive this architecture has several drawbacks. First, computing the eigendecomposition of $\bbL$ may become computationally infeasible for large graphs, and the number of trainable parameters grows with $N$. Also, the filters depend on the eigenbasis of the Laplacian and thus the parameters cannot be shared across different graphs, which limits its usage in an inductive setting. Finally, without a smoothness constraint in the frequency response one obtains filters that are not localized in the vertex domain [cf. the smooth polynomial frequency response in \eqref{E:graphconv_spectral}, resulting in a $K$-hop localized graph filter \eqref{E:graphconv}]. To overcome these limitations, the ChebNet was proposed in~\cite{defferrard2016convolutional} by defining a filter in terms of Chebyshev polynomials of the diagonal matrix of eigenvalues $\bbLambda$. In terms of expressive power, this choice is essentially equivalent to the polynomial graph filters \eqref{E:graphconv}; see also~\cite{gama2018convolutional,gama2020spmag}.

Adoption of first-order ($K=1$) graph convolutional filters was advocated for the GCN model in~\cite{kipf2016semi}, working with the degree-normalized Laplacian $\bbD^{-1/2}\bbL\bbD^{-1/2}$ as GSO and letting $\theta=h_0/2=-h_1$ so that \eqref{E:graphconv} simplifies to 
\begin{equation}\label{E:1st2}
	\bbH\bbx = \theta(\bbI_N + \bbD^{-1/2}\bbA\bbD^{-1/2})\bbx.
\end{equation}
This motivates a simple per-layer filtering update implemented to refine the nodal embeddings, namely
\begin{equation}\label{E:gv}
\bbX^{(\ell)} = \textrm{ReLU}\left(\tilde{\bbA}\bbX^{(\ell-1)} \bbTheta^{(\ell)}\right) ,
\end{equation}
where $\tilde{\bbA} := \bbI_N + \bbD^{-1/2}\bbA\bbD^{-1/2}$ is a GSO, $\bbX^{(\ell)}\in \reals^{N\times d_{\ell}}$ are the nodal representations at layer $\ell$, and $\bbTheta^{(\ell)}\in \reals^{d_{\ell-1}\times d_{\ell}}$ stores the learnable parameters of $d_{\ell}$ filters acting on $d_{\ell-1}$ input features. Notice that actually~\cite{kipf2016semi} advocates a slightly different GSO than $\tbA$, which is obtained from $\bbA$ using a renormalization trick that we describe in Section \ref{sec:method}. Going back to the general MPNN model in Section \ref{ssec:message_passing}, GCNs implement the $\text{AGGREGATE}$ function through multiplication with $\bbD^{-1/2}\bbA\bbD^{-1/2}$ and the $\text{COMBINE}$ function via: (i) the introduction of self-loops in $\bbI_N$; (ii) weighting with $\bbTheta^{(\ell)}$; followed by (iii) a point-wise nonlinear activation $\textrm{ReLU}(x) = \max(0,x)$.  

In the sequel, for the proposed encoder architecture we leverage GCN layers in~\eqref{E:gv} due to their simplicity and satisfactory performance on weighted graphs~\cite{kipf2016semi, xu2018powerful,you2020design}, such as brain SC.  While using only first-order filters, GCNs have the capability to integrate multi-hop information within $G$ using $L>1$ stacked layers. Adoption of more expressive filters where $K>1$ is left as future work, and may be prudent for larger neuroimaging datasets than the one described next.


\section{Problem setup and proposed method}\label{sec:method}

Given brain SC graphs, we build and train a model with the twofold goal of: (i) reconstructing the corresponding FC networks from nodal embeddings;  and (ii) classifying a cohort of subjects using graph-level representations learnt from binary labels in a supervised fashion. This way, learnt representations should capture both population patterns and subject-level variability. We start by describing the dataset in Section \ref{ssec:dataset}. The proposed supervised graph encoder-decoder model is then presented in Section \ref{ssec:model}. 

\begin{figure}[!t]
\centering
\subfloat[SC]{\includegraphics[scale=0.5]{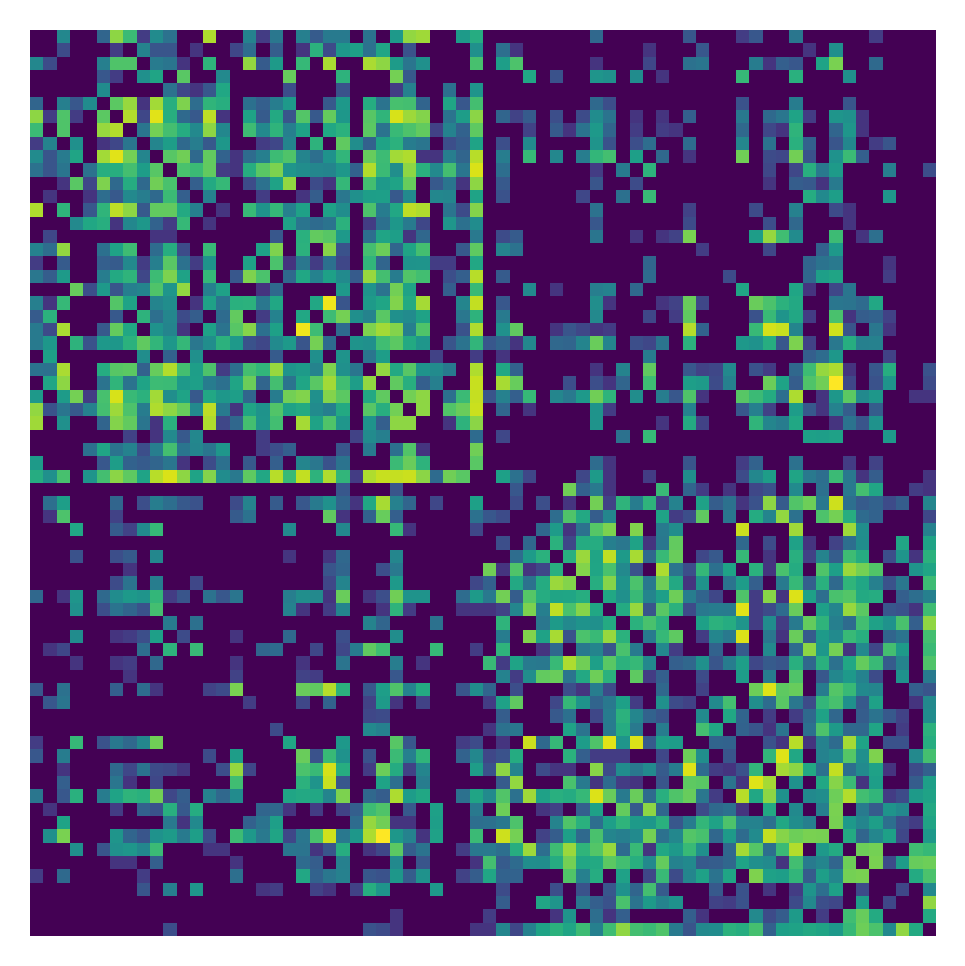}%
\label{fig_SC}}
\hfil
\subfloat[FC]{\includegraphics[scale=0.5]{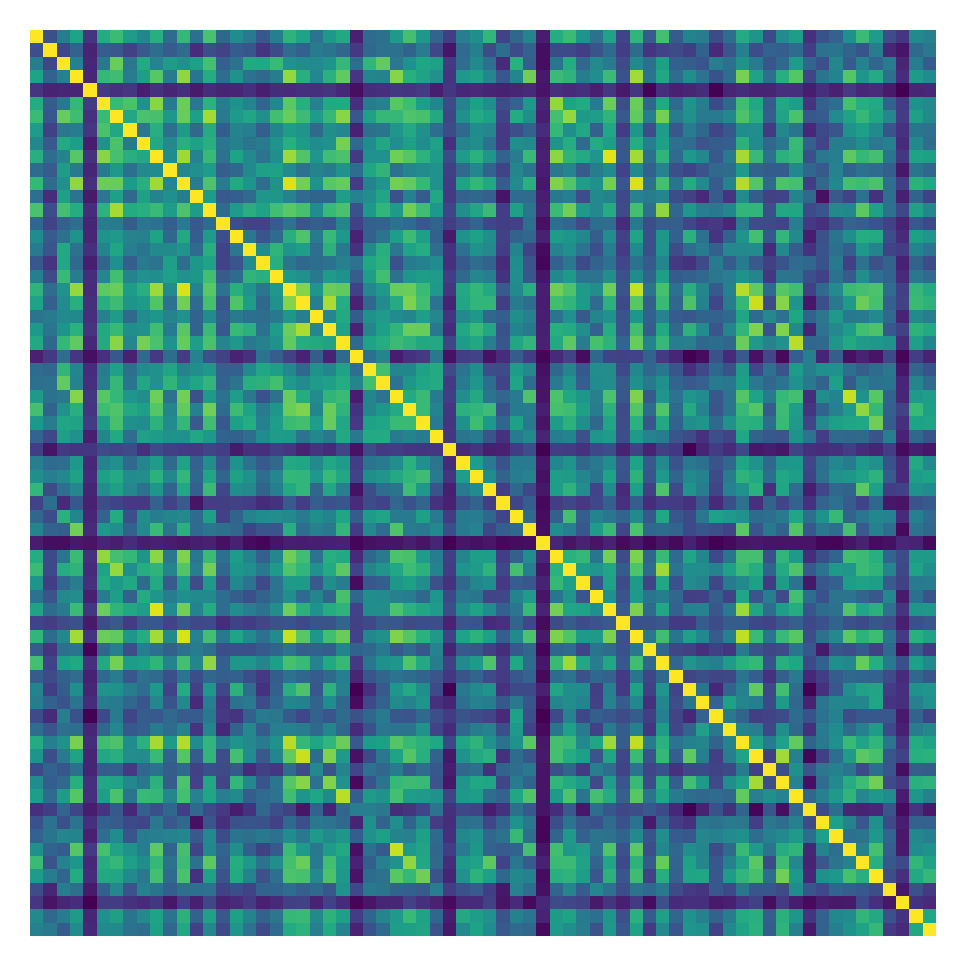}%
\label{fig_FC}}
\caption{Visualization of SC and FC networks of a sample subject, where RoIs have been ordered according to the hemisphere they belong to. Note the increased edge density in the main diagonal blocks of the SC adjacency matrix, corresponding to a modular structure with more connections within hemispheres than between. The FC graph exhibits a more disassortative pattern of connections among brain RoIs. Increased functional correlations between regions that belong to different hemispheres are observed.}
\label{fig:fig_FCSC}
\end{figure}

\subsection{Data description and network construction}\label{ssec:dataset}

We adopt a neuroimaging dataset with $P=412$ subjects from the Human Connectome Project (HCP)~\cite{van2013wu,glasser2016human}. The cohort is partitioned into two classes: 191 non-drinkers and 221 heavy drinkers according to information available on the lifetime maximum number of drinks had in a single day. The threshold is set to 21. The Desikan-Killiany atlas is used to specify brain RoIs~\cite{desikan2006automated}. Hence, $\ccalV$ in both FC and SC networks correspond to $N=68$ cortical surface regions, with 34 nodes in each hemisphere. Based on the data-processing pipeline in~\cite{zhang2019tensor,zhang2018mapping}, the SC network $\bbA$ of each subject is extracted from the dMRI and structural MRI data. Brain functional activities on each RoI are given by the resting-state BOLD time courses measured using fMRI~\cite{richiardi2013machine}. Brain FC networks $\bbSigma$ are then constructed so that edge weights are the Pearson correlation coefficient between the BOLD signals at the incident RoIs. We find that the resulting FC graphs contain few negative edge weights with much smaller magnitude compared to a vast majority of positive edges. To mitigate such data imbalance problem, we discard all functional edges with negative weights and restrict ourselves to entries $\Sigma_{ij}\in[0,1]$, as it is customarily done in prior FC studies~\cite{power2010development,rubinov2010complex}. For additional details about the data, preprocessing, and network construction steps, refer to~\cite{zhang2019tensor,zhang2018mapping} and \texttt{\url{http://www.humanconnectome.org/}}.

\begin{figure*}[!t]
\centering
\includegraphics[height = 40mm, width=\textwidth]{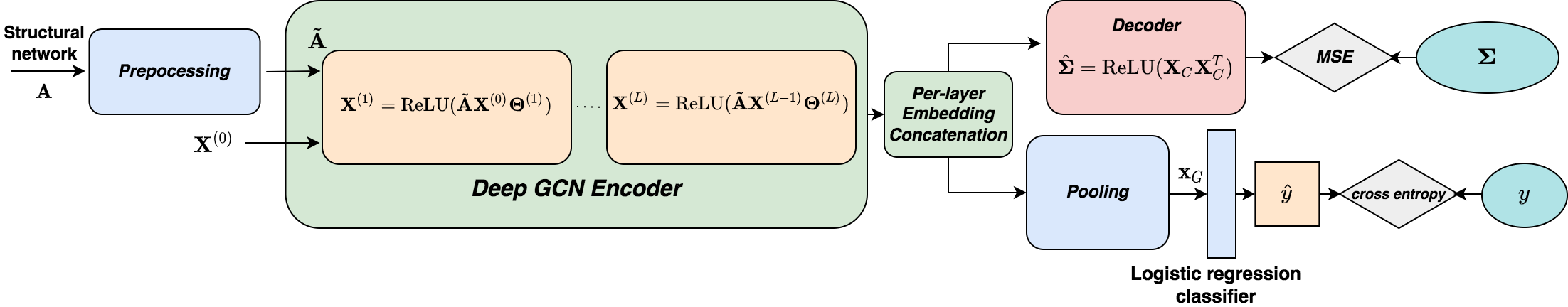}
\caption{The GRL encoder-decoder architecture. The inputs are SC networks $\bbA$ and optional nodal attributes $\bbX^{(0)}$. Graph convolution/information propagation occurs within the deep GCN encoder. Rows in $\bbX^{(L)}$ are low-dimensional node embeddings at the final layer of the GCN encoder. Outer-product decoder reconstructs the FC network $\hbSigma$ and thus implicitly models the SC-to-FC mapping. Given graph (i.e., subject)-level representations $\bbx_G$ obtained via pooling, a logistic regression classifier outputs the predicted binary label $\hat{y}$. Training is performed over a large cohort of HCP subjects, using empirical FC $\bbSigma$ and labels $y$ to define a judicious loss function.}
\label{fig:scheme}
\end{figure*} 
In Figure~\ref{fig:fig_FCSC}, we depict the resulting adjacency matrices of the SC and FC networks of a sample subject in the dataset. Nodes are rearranged by hemisphere, with the first $34$ vertices corresponding to RoIs that belong to the left brain. Note how edges densely concentrate in the main diagonal blocks of the SC network, thus indicating a markedly larger amount of anatomical connections within hemispheres rather than across the left and right brains. On the other hand, the FC network exhibits a more disassortative (or less modular) connectivity pattern. This supports the well-documented finding that functional links exist between RoIs that are loosely-connected in a structural sense, thus also depending on indirect (multi-hop) anatomical pathways. These general observations are consistent across subjects. 

\subsection{Graph encoder-decoder model and architecture}\label{ssec:model}

Here we introduce the proposed graph encoder-decoder system used to model the relationship between structural and functional brain connectomes; see also the schematic description of the architecture in Figure~\ref{fig:scheme} and the recent survey~\cite{chami2020machine}. The top branch in Figure~\ref{fig:scheme} implements a regression model to reconstruct FC from SC. The bottom branch is a supervised graph classification algorithm to predict heavy-drinking subjects. Notice that the encoder module is shared and trained using HCP data to accomplish both tasks. This way, the learnt representations capture population patterns in the SC-FC relationship while also reflect subject-specific variability to discriminate among classes. 

Next, we introduce the main modules of the architecture and explain various design choices made.

\noindent\textbf{Input.} The required input for the GRL model is an SC network represented by its (symmetric) adjacency matrix $\bbA\in\mbR^{N\times N}$, where $N=68$ is the number of RoIs from the Desikan atlas. Optionally, nodal attributes or graph signals $\bbX \in \reals^{N \times d_0}$ can be fed to the model to provide additional information via RoI-specific feature vectors of length $d_0$. Since we did not extract RoI attributes from the dataset outlined in Section \ref{ssec:dataset}, we use one-hot encoding to define graph signals and accordingly set the initial signal input as $\bbX^{(0)} = \bbI_{68}$ $(d_0 = 68)$. Future work will be devoted to collect and investigate the merits of incorporating additional meaningful subject-related nodal attributes, in addition to the graph structure itself. Examples of potential graph signals include RoI volume, node degree in the SC network, or BOLD timecourses. The latter two alternatives have been considered in network-based machine learning (ML) studies of neuroimaging data~\cite{goni2014resting,huang2018graph,parisot2018disease}. 

For training, empirical functional networks $\bbSigma$ and subject (i.e., graph-level) labels $y$ related to drinking are used to define a loss function we describe later in this section. Accordingly, both these quantities could be viewed as inputs to the model during the training phase. 

\noindent\textbf{Encoder.} The deep encoder module is a parametric function
\begin{equation}\label{E:encoder}
    \bbX_C = \textrm{ENC}(\tbA,\bbX^{(0)};\bbTheta_E)
\end{equation}
that takes a normalized version of the SC adjacency matrix $\tbA$ as well as graph signals $\bbX^{(0)}$ as inputs, and generates low-dimensional representations for each node that we stack as rows of $\bbX_C$. Following the renormalization idea in~\cite{kipf2016semi}, the normalized adjacency matrix is $\tbA := \hat{\bbD}^{-1/2}\hat{\bbA}\hat{\bbD}^{-1/2}$, where $\hbA := \bbI_N + \bbA$ and $\hbD=\textrm{diag}(\hbA\mathbf{1})$ is the degree matrix of $\hbA$. Among various possible node embedding approaches, neighborhood-aggregation methods such as the GCN model discussed in Section \ref{ssec:gcn} are simple, permutation-invariant, and inductive~\cite{bai2019unsupervised}. As a result, we opted for a GCN-based encoder to generate latent variables capturing network topology information such as the connection strengths among RoIs in SC networks. 

Unlike the single-layer GCN model in our previous work\cite{li2020supervised}, here we consider a deep GCN encoder with $L\geq 1$ stacked layers and multiple filters $d_\ell$ per layer $\ell$ as listed in Table~\ref{table_gcn}. The methodology used to choose the values of the hyperparameters $L$ and $\{d_\ell\}$ is outlined in Section \ref{ssec:training}. Another alternative to gain in expressivity is to utilize graph convolutional filters \eqref{E:graphconv} with $K>1$. But given the limited number of training samples and the satisfactory performance obtained with first-order filters, we decided to stick with GCN architecture. Recall from \eqref{E:gv} that the weight matrix $\bbTheta^{(\ell)} \in\reals^{d_{\ell-1} \times {d_\ell}}$ contains learnable graph filter coefficients at layer $\ell$. So overall, the trainable parameters in the deep GCN-based encoder \eqref{E:encoder} are $\bbTheta_E:=\{\bbTheta^{(\ell)}\}_{\ell=1}^L$.  The {ReLU} activation function is used to speed up training and avoid the problem of vanishing gradients in multi-layer settings~\cite{maas2013rectifier}. 

It is natural to consider the output $\bbX^{(L)}\in\reals^{N\times d_L}$ of layer $L$ as the sought nodal embeddings. Inspired by~\cite{xu2018powerful,bai2019unsupervised}, we also evaluated the contribution from node embeddings learnt in intermediate GCN layers $\ell=1,\ldots,L-1$ via concatenation, thus forming $\bbX_C\in\reals^{N\times \sum_{\ell}d_\ell}$.

\noindent\textbf{Decoder and loss function.} Starting with the FC reconstruction branch of the GRL architecture in Figure \ref{fig:scheme}, node embeddings $\bbX_{C}$ go through an outer product decoder
\begin{equation}\label{E:decoder}
	\hbSigma = \textrm{ReLU}(\bbX_{C}\bbX_{C}^{\top})
\end{equation}
to generate a predicted FC adjacency matrix $\hbSigma$. Notice that we utilize a rectifier to enforce the constraint $\Sigma_{ij}\geq 0$. The mean squared error (MSE) between the reconstructed graph $\hbSigma$ and the empirical FC $\bbSigma$, averaged over a training set and denoted as $\mathcal{L}_\textbf{MSE}({\hbSigma}, \bbSigma)$, is used as reconstruction loss for training the regression branch of the model.

Moving on to the supervised classification branch of the model, we compute a graph-level embedding by taking the row-wise average of all nodal representations in $\bbX_C$. This results in a vector $\bbx_G$ that summarizes information about the SC-FC relationship [cf. \eqref{E:gpool}]. Admittedly simple, average pooling has been shown effective in multiple studies~\cite{dai2016discriminative,duvenaud2015convolutional}. For a thorough comparison, we also investigated max pooling and sum pooling which are common alternatives~\cite{hamilton2017inductive}. The graph embeddings $\bbx_G$ are then fed to a logistic regression classifier to predict the binary subject labels. As our main objective is to assess how well the graph convolution operation helps integrate information within brain networks, a simple logistic regression classifier is preferred rather than advanced models such as multi-layer perceptrons, where training may be biased towards the fully connected layers post GCN. 

Finally, the GRL model is trained on HCP data in an end-to-end manner by considering the regression and classification tasks simultaneously. The overall loss function is given as
\begin{equation}\label{E:loss_func}
	\ccalL = \ccalL_\textbf{MSE}(\hbSigma, \bbSigma) + \lambda \times \ccalL_\textbf{CLA}(\hat{y},y),
\end{equation}
where $\hat{y}$ is the predicted label and $y$ is the ground truth label of the subject. The sigmoid cross-entropy loss denoted by $\mathcal{L}_\textbf{CLA}$ is used to evaluate classification performance.  Hyperparameter $\lambda$ controls the trade-off between FC reconstruction and classification performance. Its value is determined via grid search and predefined criteria described in the next section. By training the model end-to-end with the loss function as in~\eqref{E:loss_func}, we aim to strike the right balance between FC reconstruction and subject classification and achieve satisfactory performance on both objectives. The results are presented in the Section \ref{sec:results}.


\section{Training and Evaluation Framework}\label{sec:training_evaluation}

Training details and hyperparameter search criteria are outlined in Section \ref{ssec:training}; the code to implement the model is publicly accessible from \texttt{\url{https://github.com/yli131/brainGRL/}}. We also describe our comprehensive evaluation protocol in Section \ref{ssec:evaluation}, listing several baselines that will guide the presentation of numerical results in Section \ref{sec:results}.

\subsection{Training details}\label{ssec:training}

The supervised graph encoder-decoder system is implemented in TensorFlow\cite{abadi2016tensorflow}. The details about the model architectures we tested are given in Table~\ref{table_gcn}. We carry out a two-stage training procedure for model selection. 

The first stage is a low-resolution model selection module. We set $\lambda = 0.1$ following our preliminary results in~\cite{li2020supervised}, and search for the optimal architecture among the options in Table~\ref{table_gcn}. With fixed $\lambda$, during training of each candidate model architecture we use $10$-fold cross validation whereby the whole dataset of $P=412$ subjects is randomly partitioned into $80\%$ training, $10\%$ validation, and $10\%$ test set. The random seed for each fold is fixed for a fair comparison between model architectures. We use Xavier initialization for the weight coefficients $\bbTheta^{(\ell)}$ of each GCN layer $\ell$~\cite{glorot2010understanding}. The Adam algorithm with learning rate $0.001$ was adopted to optimize the tunable parameters in $\bbTheta_E$ as well as those of the logistic regression classifier~\cite{kingma2014adam}. To avoid overfitting, early stopping is applied to monitor the validation loss and to stop the training once the loss increases during $10$ consecutive training epochs. Altogether, we trained $57$ models because we also considered three global pooling mechanisms (mean/max/sum pooling) and node embedding concatenation for the multi-layer GCN encoders (with $L=2$ or $L=3$).

\begin{table}[!t]
\renewcommand{\arraystretch}{1.3}
\caption{GCN architectures for hyperparameter search}
\label{table_gcn}
\centering
\begin{tabular}{|c|c|}
\hline
$L=1$ layer & $d_0 \times \{128; 64; 32; 16; 8\}$\\
\hline
$L=2$ layers & $d_0 \times \{128 \times 64; 64 \times 32; 32 \times 16; 16 \times 8\}$\\
\hline
$L=3$ layers & $d_0 \times \{128 \times 64 \times 32; 64 \times 32 \times 16; 32 \times 16 \times 8\}$\\
\hline
\end{tabular}
\end{table}

For each of the $57$ model architectures, metrics including the graph reconstruction error, classification accuracy and F score are monitored across all $10$ folds to evaluate the performance on both regression and classification. Training time is also taken into consideration to offer insights into the performance cost ratio. A selection criterion $\ccalC_{\textrm{low}}$ is developed by combining the aforementioned metrics, which is given by 
\begin{equation}\label{E:cl}
\ccalC_{\textrm{low}} = \frac{\text{classification accuracy} \times \text{F score}}{\text{MSE} \times \text{training time}}.
\end{equation}
The three models with highest $\ccalC_{\textrm{low}}$ are singled out for the second stage. Therein, an extensive grid search over $\lambda$ values from $0$ to $1.5$ with step size $0.1$ is carried out to determine the optimal $\lambda$ for each selected GCN architecture.

Recall the loss function \eqref{E:loss_func}. For large values of $\lambda$, the reconstruction performance deteriorates because training is biased towards the subject classification objective. When $\lambda  = 0$, the model achieves the best possible FC reconstruction while the classification loss is at its peak. As $\lambda$ increases from zero, the classification loss first decreases, then fluctuates and saturates at a certain level. This is intuitive as the graph embeddings that serve as input features for classification depend on the learnt node embeddings that are also driven by the reconstruction objective. When the reconstruction is getting worse, as expected the classification performance will also eventually degrade. The $\lambda$ search results in Section \ref{ssec:training_results} will corroborate such intuition. 

In the end, among the $3$ picked encoder architectures and all $16$ possible $\lambda$ values, we choose the final model for downstream analysis and visualization  as the one that maximizes
\begin{equation}\label{E:ch}
\ccalC_{\textrm{high}} = \frac{\text{classification accuracy} \times \text{F score}}{\text{MSE}}.
\end{equation} 

\begin{figure*}[!ht]
\centering
\subfloat[Graph reconstruction performance]{\includegraphics[height = 45mm, width=0.33\textwidth]{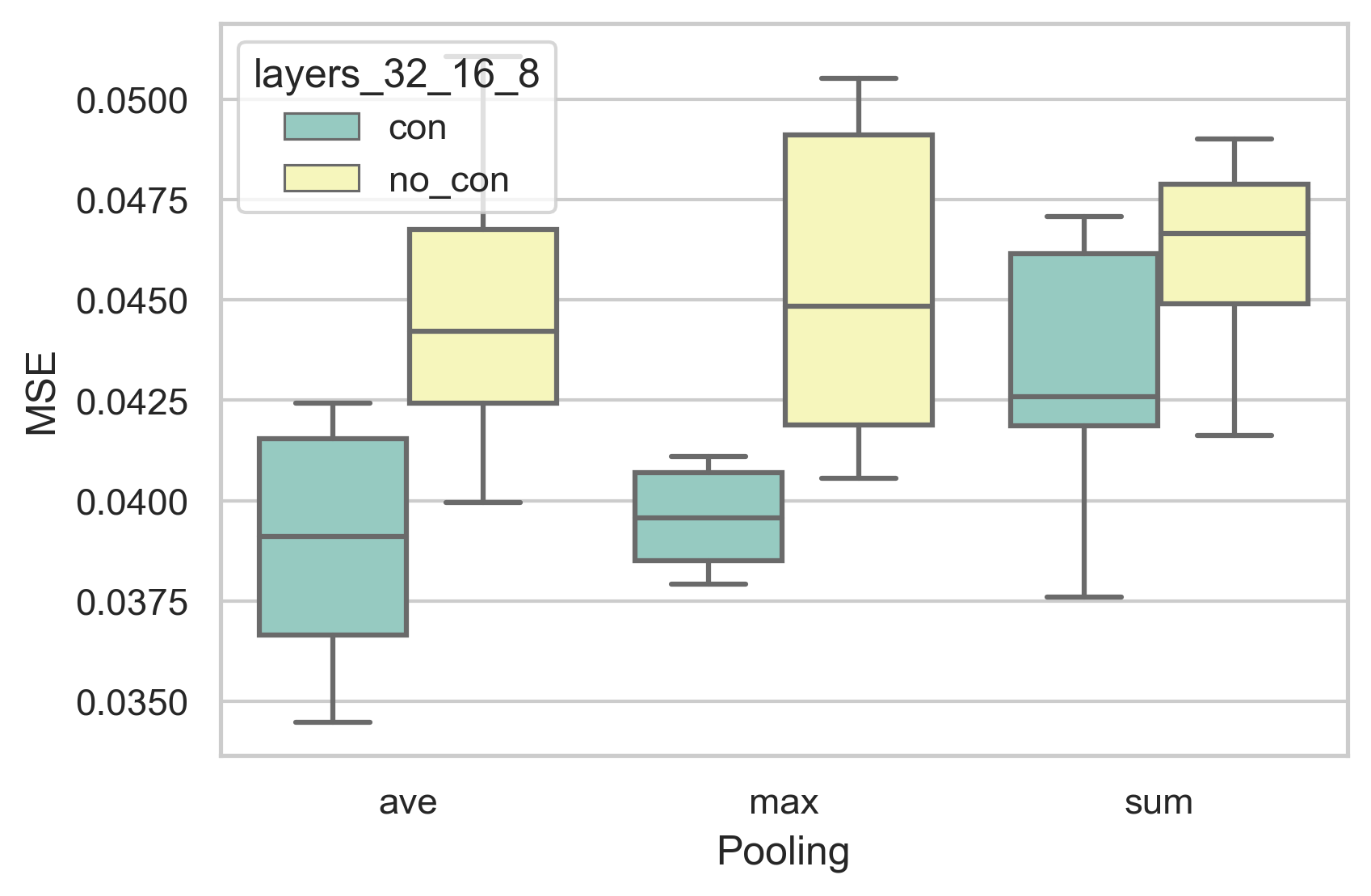}%
\includegraphics[height = 45mm, width=0.33\textwidth]{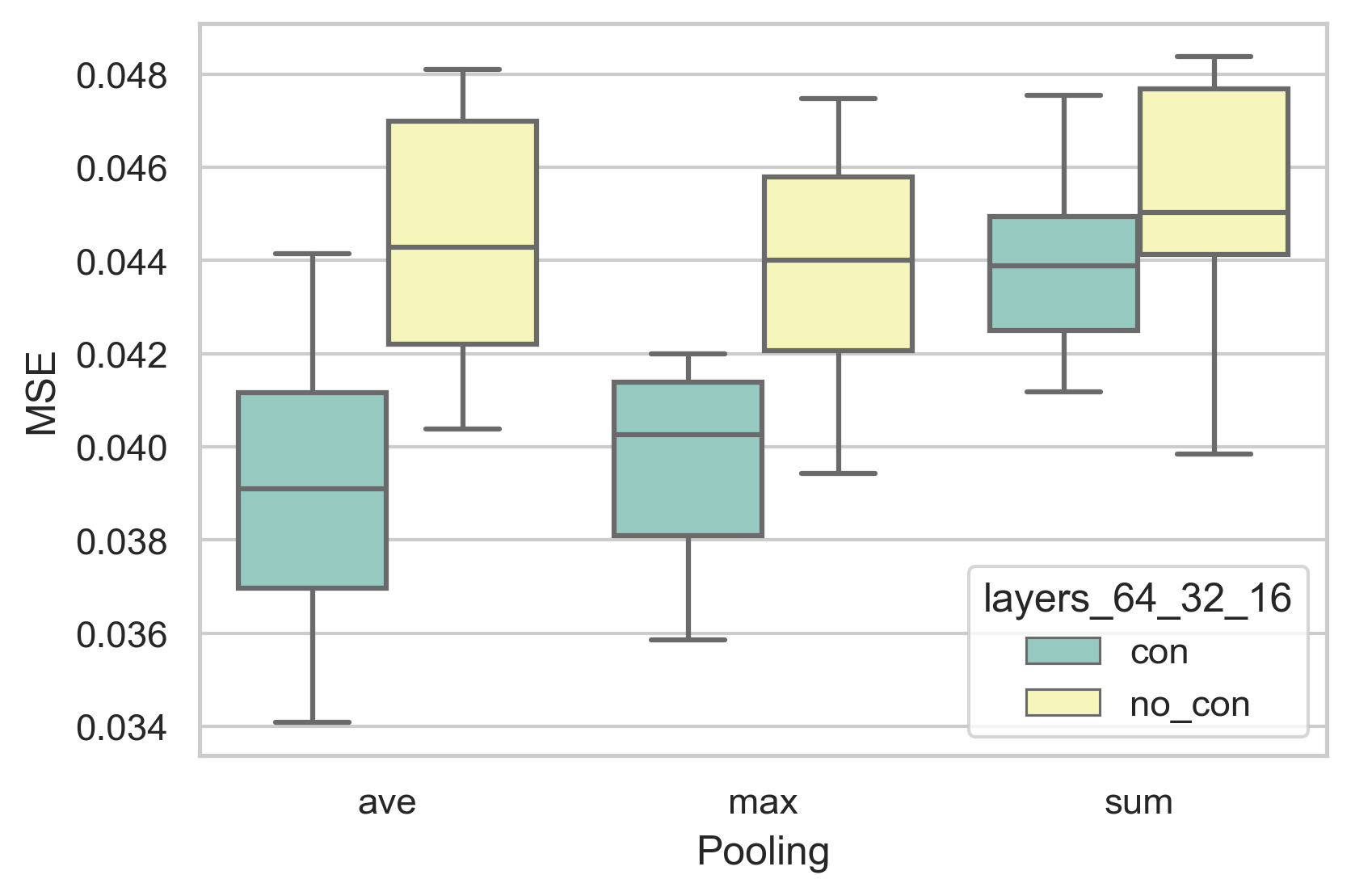}%
\includegraphics[height = 45mm, width=0.33\textwidth]{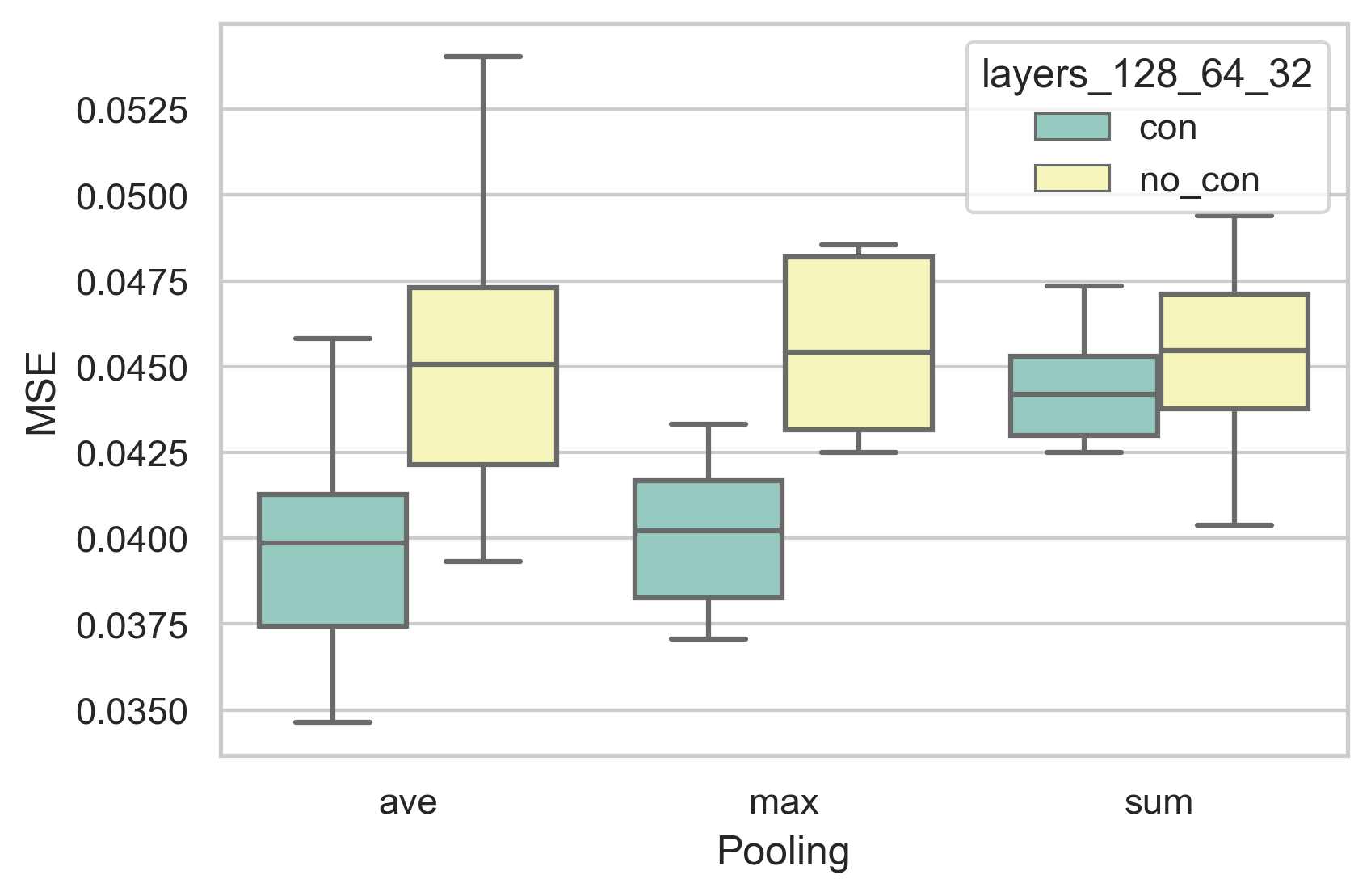}%
\label{regressionperf}}
\hfil
\subfloat[Subject classification performance]{\includegraphics[height = 45mm, width=0.33\textwidth]{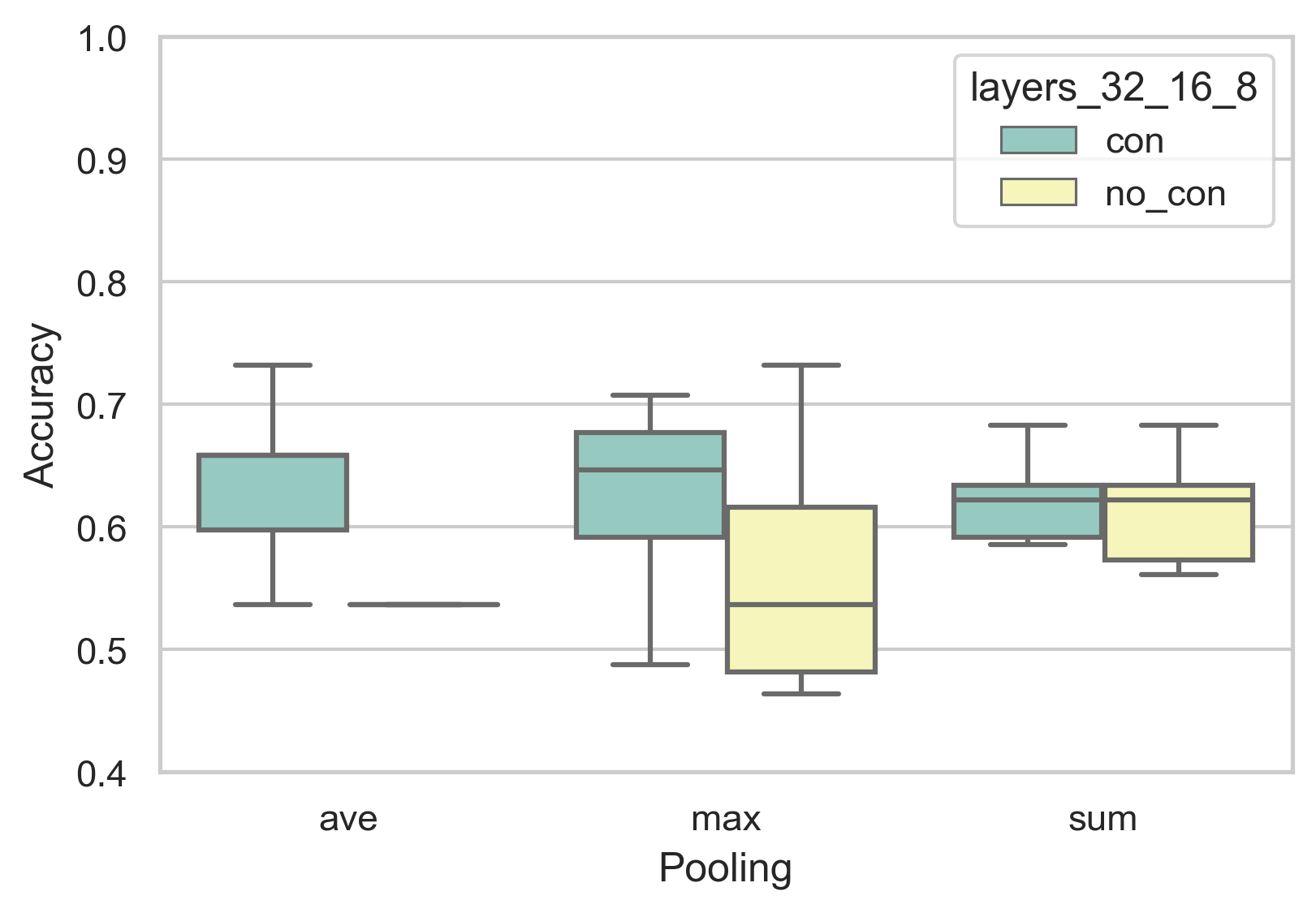}%
\includegraphics[height = 45mm, width=0.33\textwidth]{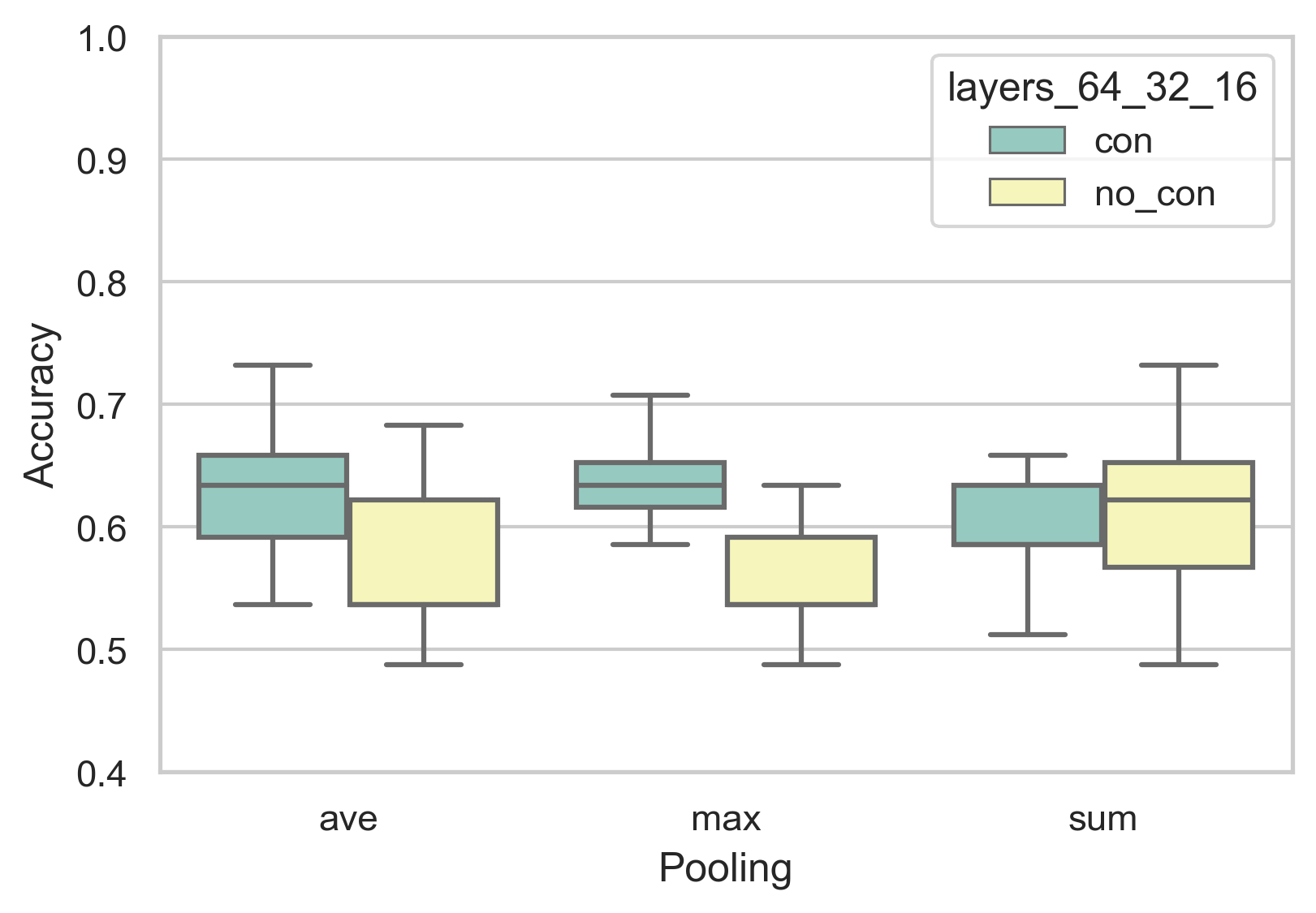}%
\includegraphics[height = 45mm, width=0.33\textwidth]{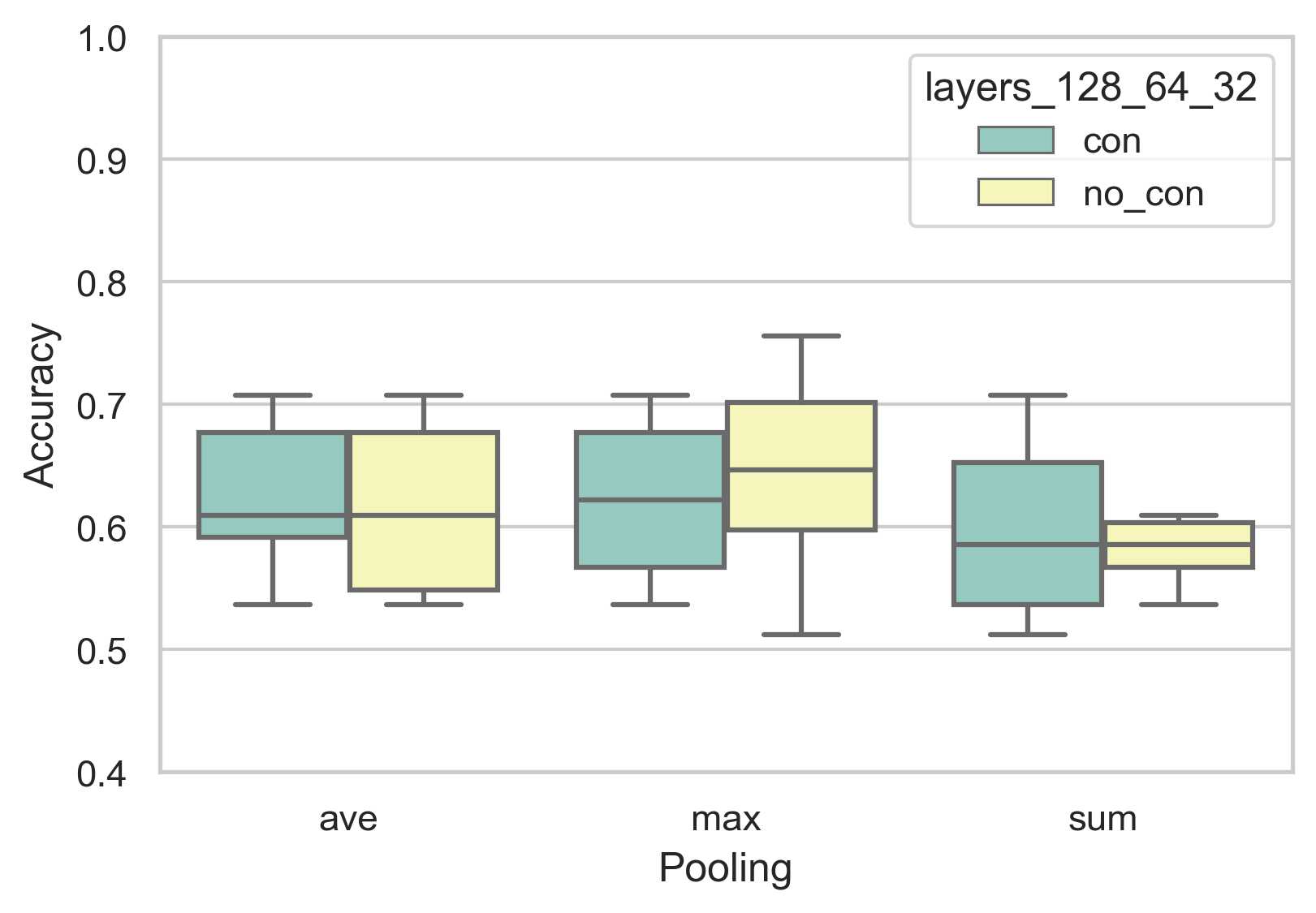}%
\label{classperf}}
\caption{Graph reconstruction and subject classification performance of the three selected model architectures with $\lambda = 0.1$ and $L=3$ (number of filters per layer shown in the legend where 'con' represents concatenation of node embeddings from each layer and 'no\_con' indicates that only the node embedding from the last layer is considered). Apparently, the combination of  node embedding concatenation and global average pooling attain better performance in graph reconstruction with lower MSE and subject classification with higher accuracy.}
\label{fig:threemodelaccmse}
\end{figure*}

\subsection{Evaluation protocol and baselines}\label{ssec:evaluation}

To corroborate the effectiveness of the proposed model, in Section \ref{ssec:baselines} we perform a comprehensive comparative study involving four baseline methods. 

First, instead of training the graph classification framework in an end-to-end manner, we train the graph reconstruction branch with $\lambda = 0$~\cite{GlobalSIP19_YangLiEncoderDecoder}. Subsequently, the graph embeddings obtained by row-wise average pooling of node embeddings are fed to a logistic regression classifier that is separately trained for subject-level classification. Two-step algorithms have been observed to yield sub-optimal performance for supervised learning application~\cite{weston2012deep}. Accordingly, we expect that training our model in an end-to-end fashion while jointly accounting for both objectives in \eqref{E:loss_func} will lead to better performance relative to this \emph{two-step} baseline.

Second, as one of the objectives of our framework is to reconstruct FC and distill information from the structure-function relationship as a byproduct, a natural alternative would be to consider using FC as an \emph{additional input} besides SC. To this end, we carry out an experiment where both the SC and FC networks of the same subject are inputs to a feed-forward GCN model, that is trained to predict the subject-level binary labels associated to drinking habits. Such comparison would help reveal whether the SC-FC mapping offers additional information for classification on top of the raw graph topologies for both modalities.

Third, we explore whether the information within SC or FC \emph{individually} suffices for subject classification. With this purpose, the scheme in Figure~\ref{fig:scheme} is modified so that the desired output graph is the same as the one fed at the input. As a result, with the input being SC or FC, the model boils down to a graph autoencoder. The training details remain the same and consequently, the learnt representations capture information with respect to either SC or FC only. Their discriminative power for subject classification is evaluated and compared against the learnt representations of the SC-FC mapping.

The last baseline method is an ML pipeline that relies on handcrafted features constructed from graph summary statistics extracted from SC, FC, or both. Network measures\cite{kolaczyk2009book,rubinov2010complex,bullmore2009complex} capture certain properties of the graph, e.g., community structure, connectivity, and graph components or subgraph structure. Here, we extracted scalar graph measures including average path length, global efficiency, clustering coefficient, graph radius/diameter, transitivity as well as graph density. We combined all aforementioned graph-level measures into a feature vector summarizing network structure. For more information regarding the said graph measures, the interested reader is referred to~\cite{rubinov2010complex,kolaczyk2009book}. After constructing such feature vector for each subject, we implement workhorse ML models such as SVM, logistic regression and tree-based classifiers; see e.g.,~\cite{elements_of_statistics}. Such comparisons with baseline ML-based methods is expected to reveal the power of the proposed GRL model, by capitalizing on task-driven learnt representations (cf. handcrafted features) that exploit the geometric structure of network data.

\begin{figure*}[!ht]
\centering
\subfloat[Reconstruction performance for lambda search]{\includegraphics[height = 45mm, width=0.33\textwidth]{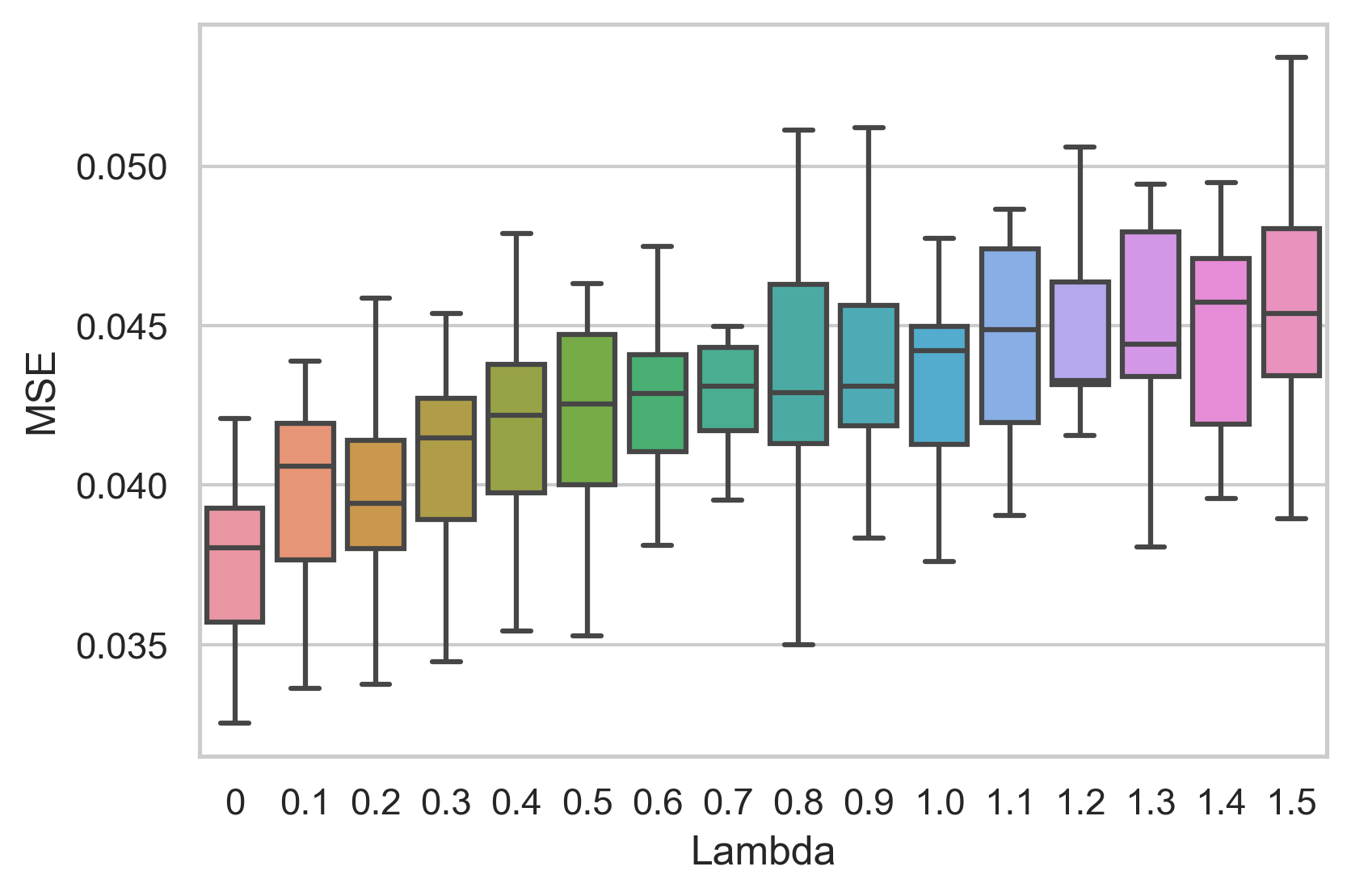}%
\includegraphics[height = 45mm, width=0.33\textwidth]{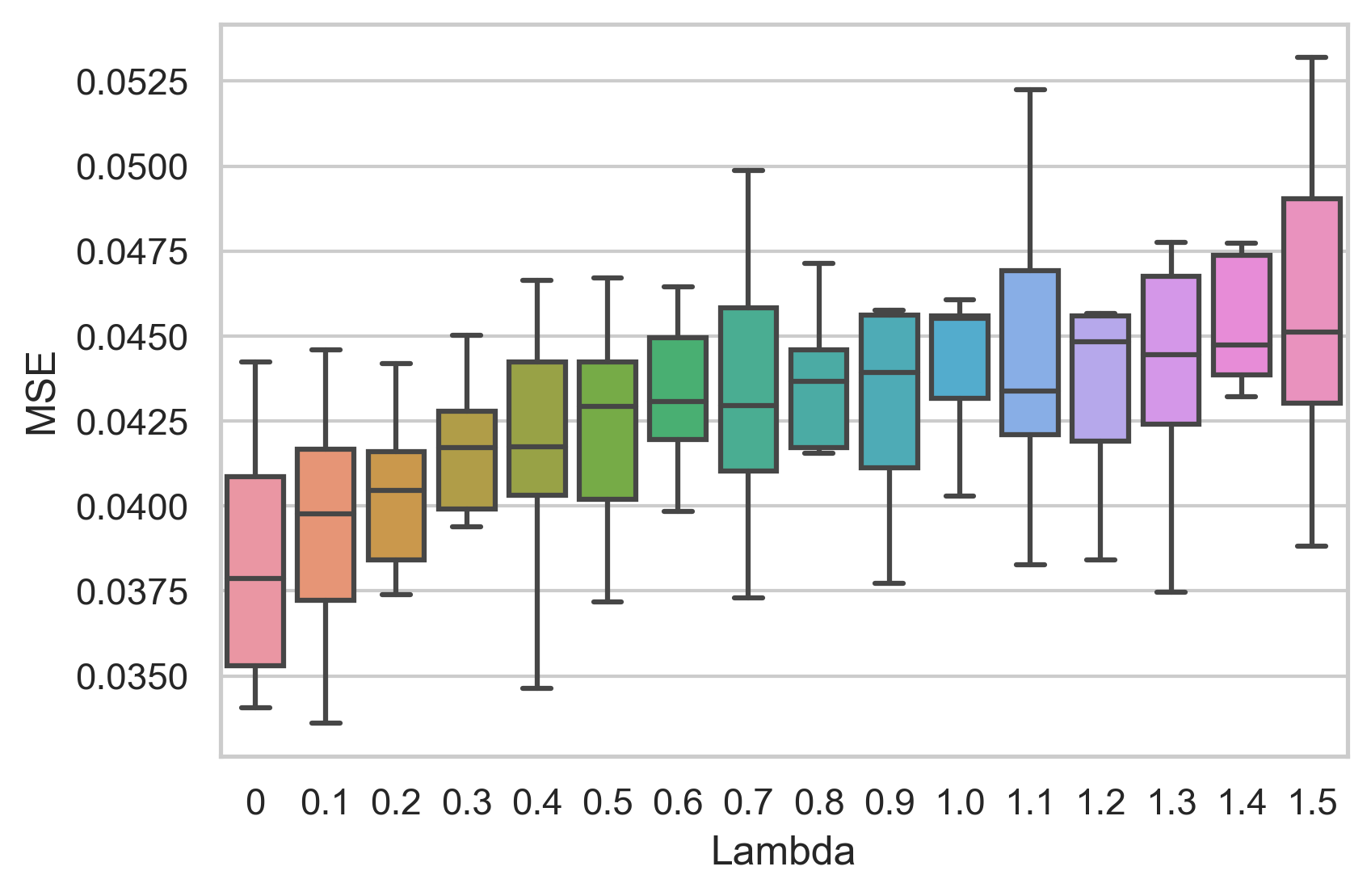}%
\includegraphics[height = 45mm, width=0.33\textwidth]{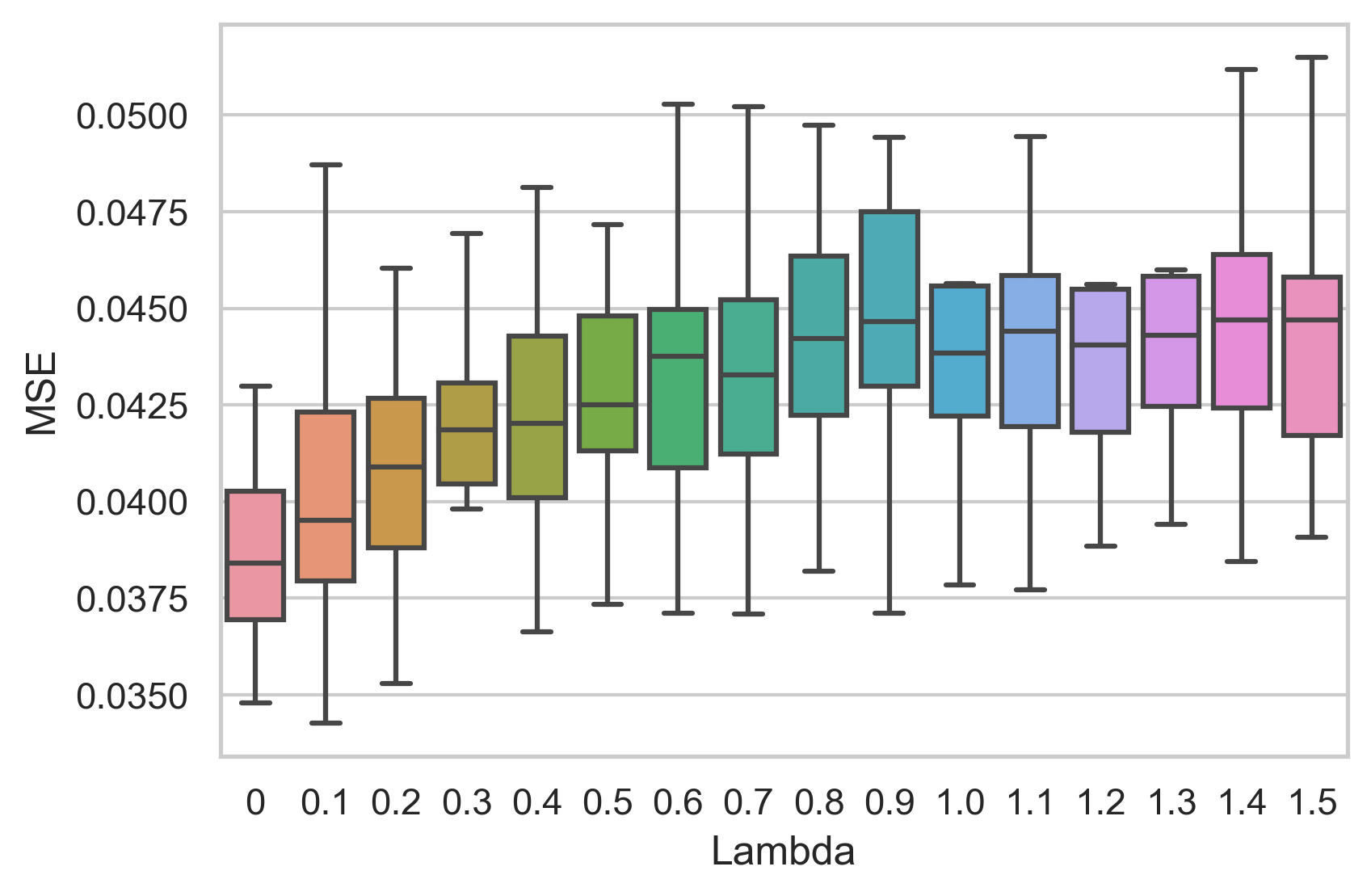}%
\label{mselambda}}
\hfil
\subfloat[Classification performance for lambda search]{\includegraphics[height = 45mm, width=0.33\textwidth]{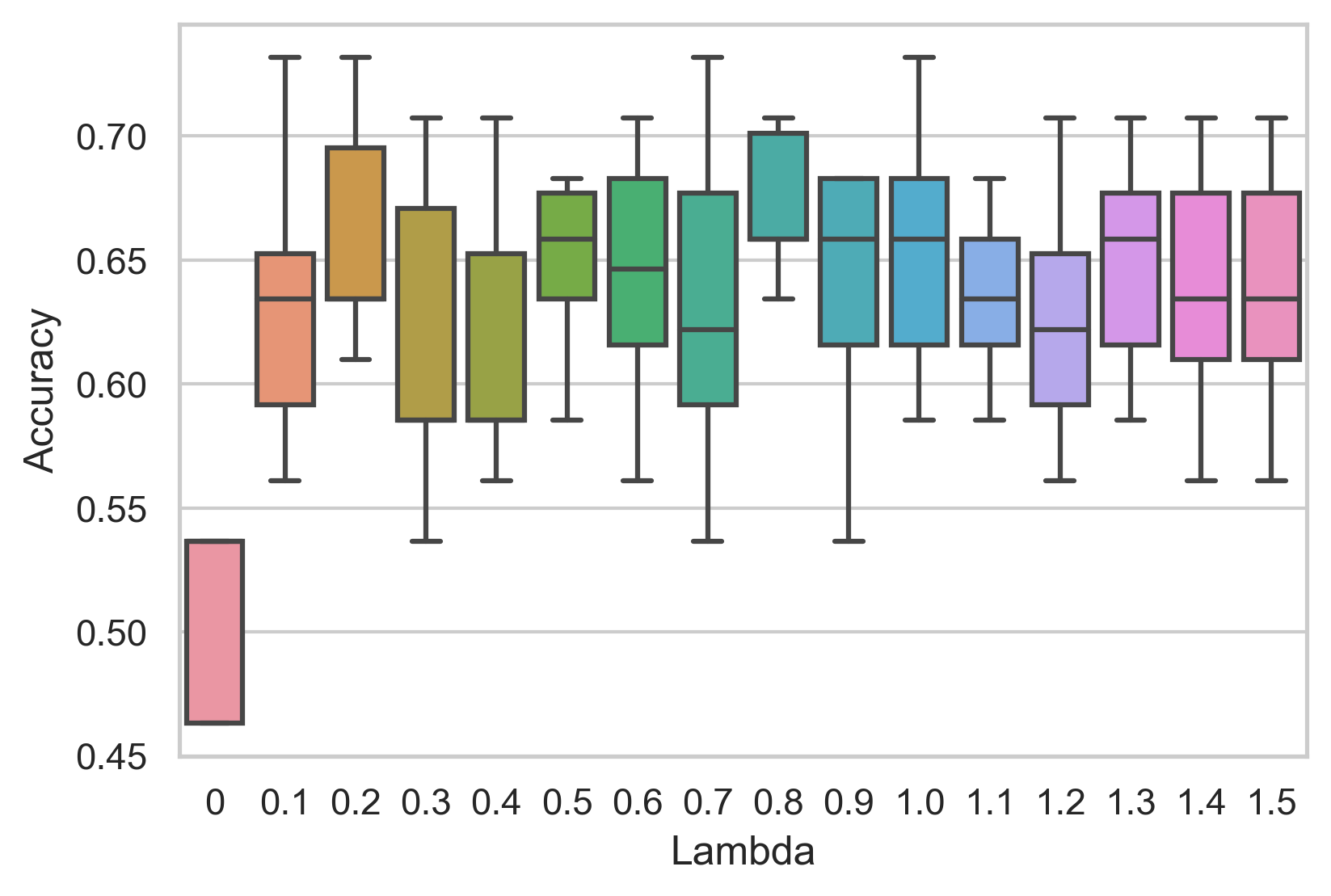}%
\includegraphics[height = 45mm, width=0.33\textwidth]{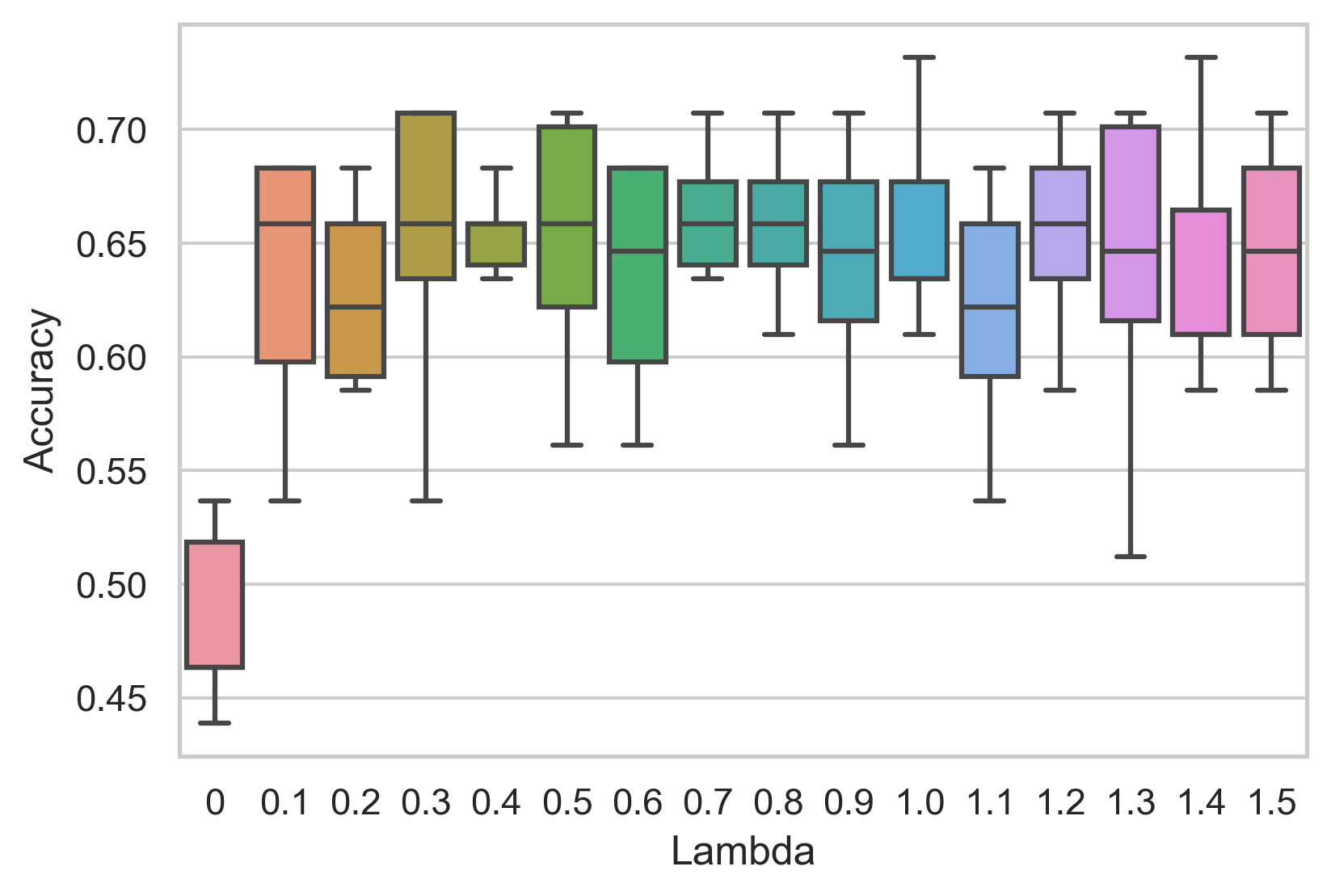}%
\includegraphics[height = 45mm, width=0.33\textwidth]{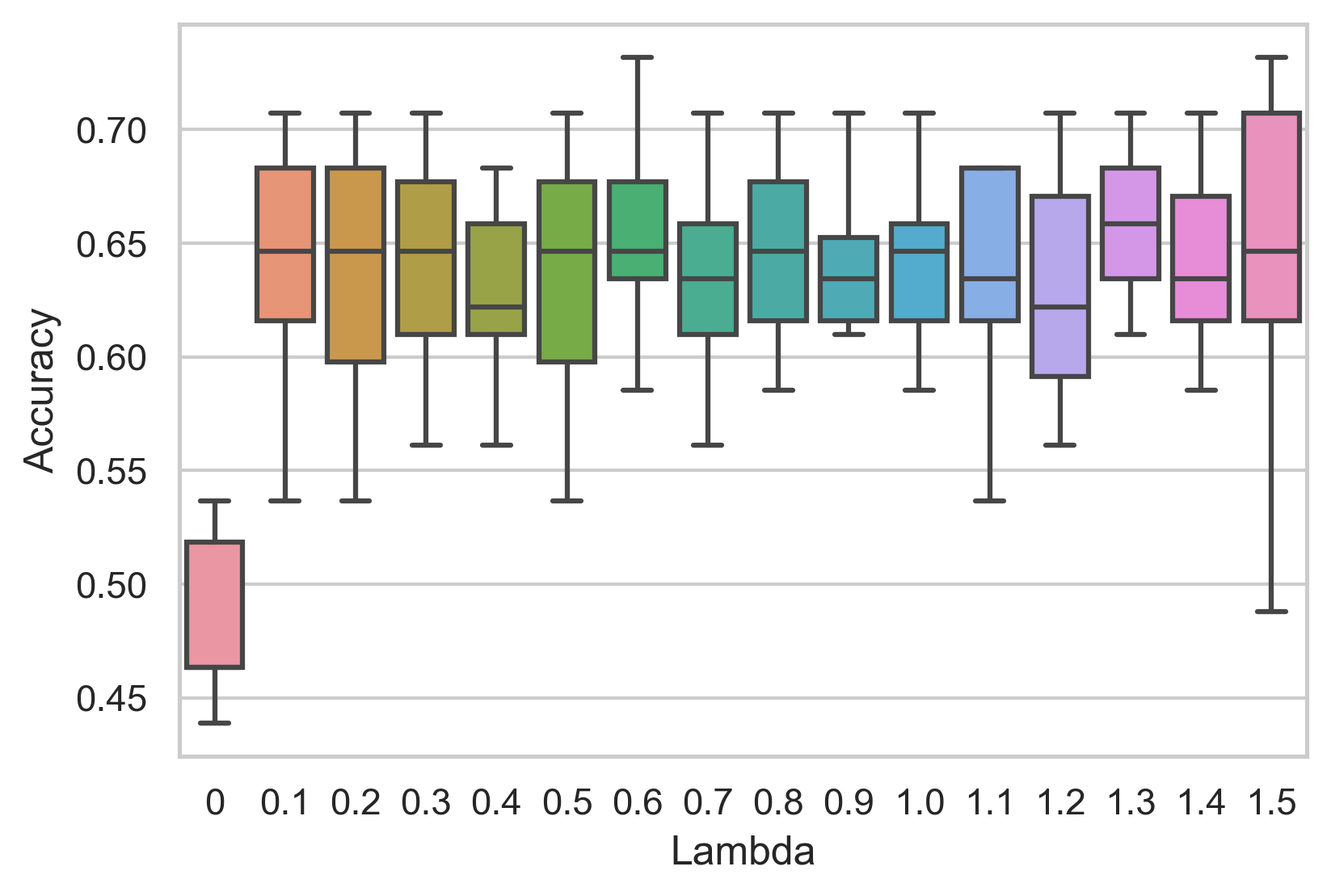}%
\label{acclambda}}
\hfil
\caption{FC reconstruction MSE and classification accuracy with respect to different $\lambda$ values, for the three selected model architectures with $32 \times 16 \times 8$, $64 \times 32 \times 16$, $128 \times 64 \times 32$ filter (left to right). As $\lambda$ increases, the training is biased towards classification thus the MSE increases. The classification performance saturates for large $\lambda$ values as the reconstruction is compromised at these values. The chosen model that maximizes $\ccalC_{\textrm{high}}$ has $L=3$ layers with $32 \times 16 \times 8$ filters (left), node embedding concatenation and global average pooling, where $\lambda=0.2$. }
\label{fig:lambdasearch}
\end{figure*}


\section{Results}\label{sec:results}

Here we present and discuss the results of our experimental validation with HCP data.

\subsection{Training and model selection}\label{ssec:training_results}

As discussed in Section \ref{ssec:training}, the first stage of training is a low-resolution model selection process, where we tested the $12$ GCN-based encoder architectures given in Table~\ref{table_gcn}. Comparisons were also conducted among global pooling methods, and between the inclusion and exclusion of node embedding concatenation in multi-layer ($L>1$) settings. Figure~\ref{fig:threemodelaccmse} depicts the results of the performance comparison for $3$-layer GCN encoders. We can tell that concatenating node embeddings from previous layers to form $\bbX_C$ helps improve the performance in both subject classification (second row) and FC graph reconstruction (first row). This finding is consistent with prior results reported in the literature~\cite{hamilton2017inductive, bai2019unsupervised}, suggesting that the embeddings in early layers also contain valuable information about localized signal propagation over the graph $G$. Based on the criteria $\ccalC_{\textrm{low}}$ defined in
\eqref{E:cl}, we singled-out the three best models. These are the models with $L=3$ GCN layers (see the third row in Table~\ref{table_gcn}), together with node embedding concatenation and global average pooling.

Starting with these three models, the second stage was an extensive grid search of the $\lambda$ parameter in \eqref{E:loss_func} to determine the optimal trade-off between graph reconstruction and subject classification. The classification accuracy and FC reconstruction MSE on the three selected models were evaluated and presented in Figure \ref{fig:lambdasearch}. It is apparent that $\lambda = 0$ gives the lowest regression MSE as the model training merely focuses on graph reconstruction. As $\lambda$ gets larger, the regression error keeps increasing while the classification accuracy slightly fluctuated around a saturation value. Finally, the best model for the downstream tasks and further analysis as well as comparison with baselines is chosen based on the criteria $\ccalC_{\textrm{high}}$ defined in \eqref{E:ch}. This corresponds to an encoder architecture with $L=3$ layers and $32 \times 16 \times 8$ filters, where $\lambda = 0.2$. Recall we concatenate nodal representations from all layers, and we use global average pooling to obtain $\bbx_{G}$. 

\subsection{Comparison with baselines}\label{ssec:baselines}
In this section, we discuss the results of the comparisons with the baselines outlined in Section \ref{ssec:evaluation}.

\noindent\textbf{Two-step GRL followed by classification procedure.} Unlike the model in Figure \ref{fig:scheme} which is trained in an end-to-end manner, GRL-based classification can also be conducted in a step-wise fashion. The first step is to train the FC reconstruction branch alone ($\lambda = 0$). In the second step, graph embeddings $\bbx_G$ are constructed from the output node representations $\bbX_C$ in the first step, and then used in a logistic regression classifier for subject classification. When it comes to training in an end-to-end or stage-wise fashion, $10$-fold cross validation is adopted for both schemes with the same random seeds for fair comparison. The performance of these two training procedures is summarized in Table~\ref{table_baseline}. As expected, the reconstruction MSE is marginally lower for the two-step procedure, as the encoder is optimized for FC reconstruction performance. When it comes to subject (i.e., graph) classification, the proposed model outperforms the two-step GRL baseline both in terms of accuracy and F score, matching the findings in~\cite{weston2012deep}. All in all, when trained in an end-to-end fashion the proposed model attains a markedly better classification performance, while incurring a marginal penalty in FC reconstruction MSE relative to a baseline that was optimized for this latter objective.

\begin{table}[!t]
\renewcommand{\arraystretch}{1.3}
\caption{End-to-end model versus two-step baseline}
\label{table_baseline}
\centering
\resizebox{\columnwidth}{!}{%
\begin{tabular}{|c|c|c|c|}
\hline
Model & MSE & Accuracy & F score\\
\hline
End-to-end & $0.0398 \pm 0.003$ & $\mathbf{0.6610 \pm 0.043}$ & $\mathbf{0.6962 \pm 0.030}$\\
\hline
Two-step & $\mathbf{0.0377 \pm 0.003}$ & $0.5072 \pm 0.033$ & $0.5071 \pm 0.033$ \\
\hline
\end{tabular}}
\end{table}

\begin{figure}[!t]
\centering
\subfloat[Proposed method]{\includegraphics[height = 40mm, width=0.5\columnwidth]{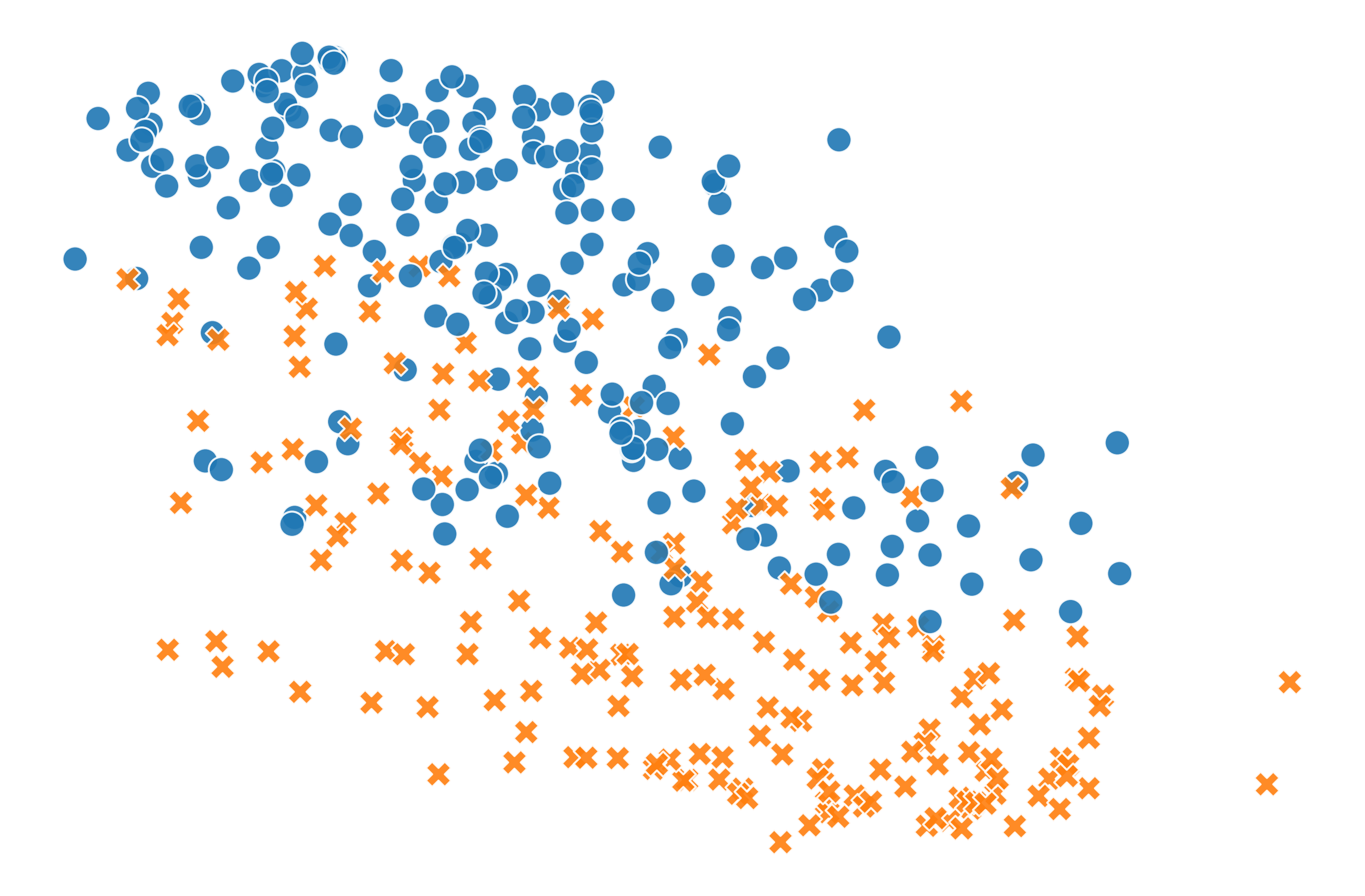}%
\label{fig:besttsne}}
\hfil
\subfloat[Two-step method]{\includegraphics[height = 40mm, width=0.5\columnwidth]{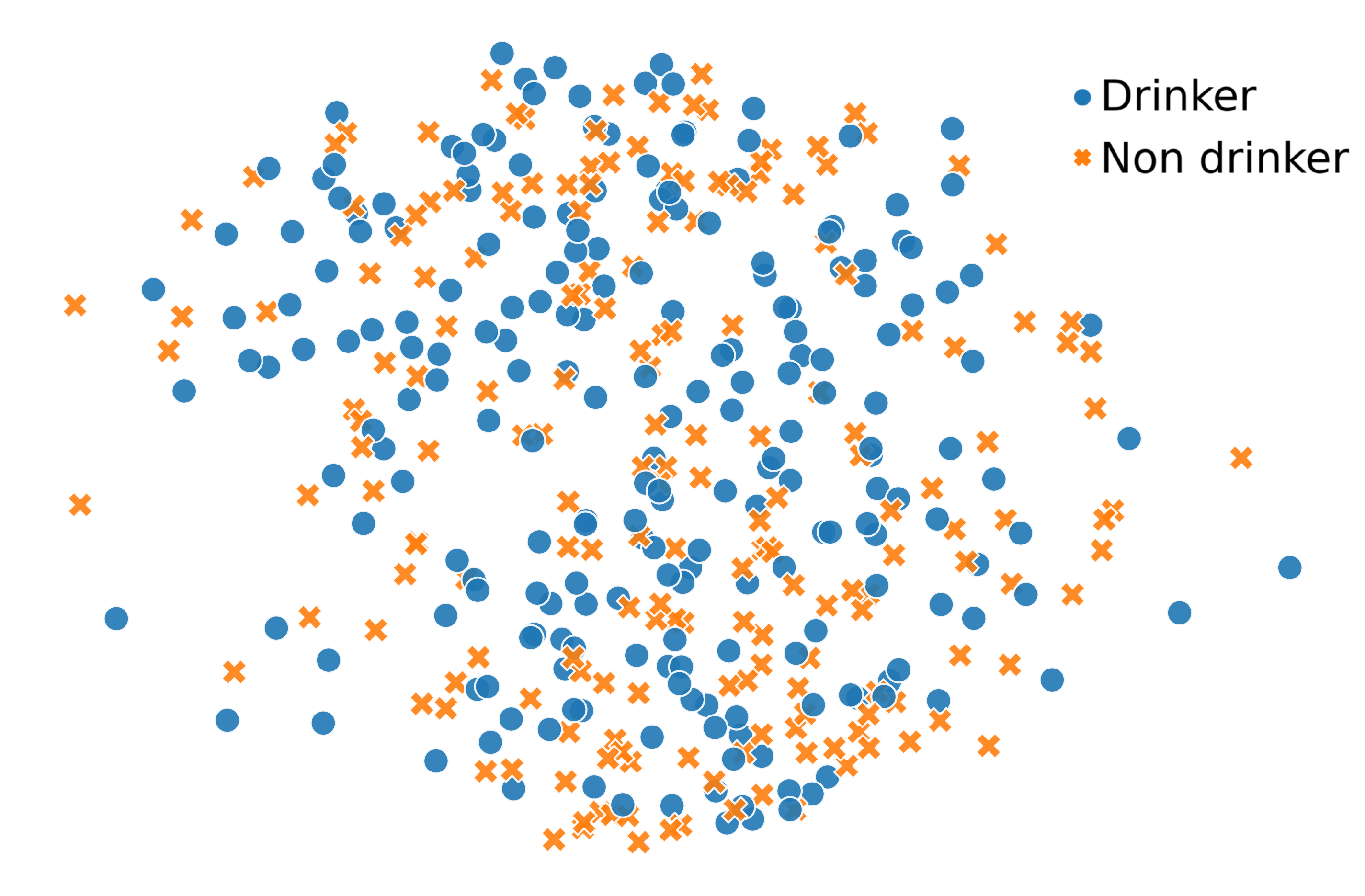}%
\label{fig:basetsne}}
\caption{2D visualization of the graph embeddings learnt by the proposed model (left) and the two-step baseline method (right). Group-wise separation can be observed on the left but not on the right, supporting the difference in classification performance reported in Table~\ref{table_baseline}.}
\label{fig:tsne}
\end{figure}

For visualization purposes, the model with the highest test accuracy among the $10$ folds is retained and re-run on the whole dataset. In more detail, the SC networks of all $P=412$ subjects are fed through the trained encoder-decoder system in Figure~\ref{fig:scheme} to yield per-subject reconstructed FC networks and the low-dimensional graph embeddings that model the SC-FC relationship. We visualize the vectors $\bbx_G$ in the 2D plane using the dimensionality reduction algorithm t-SNE\cite{maaten2008visualizing}. The results are depicted in Figure \ref{fig:besttsne}. We can distinguish a separation, albeit not perfect, between the two groups, indicating that the learnt graph representations contain valuable information for subject classification. In Figure \ref{fig:basetsne}, we also plot the learnt graph embeddings obtained when $\lambda = 0$, i.e., optimizing only the FC reconstruction loss. In that case, there is no apparent separation among the classes. These findings suggest that end-to-end training captures discriminative properties of the subjects' networks.

\begin{figure}[!t]
\centering
{\includegraphics[height = 60mm, width=0.9\columnwidth]{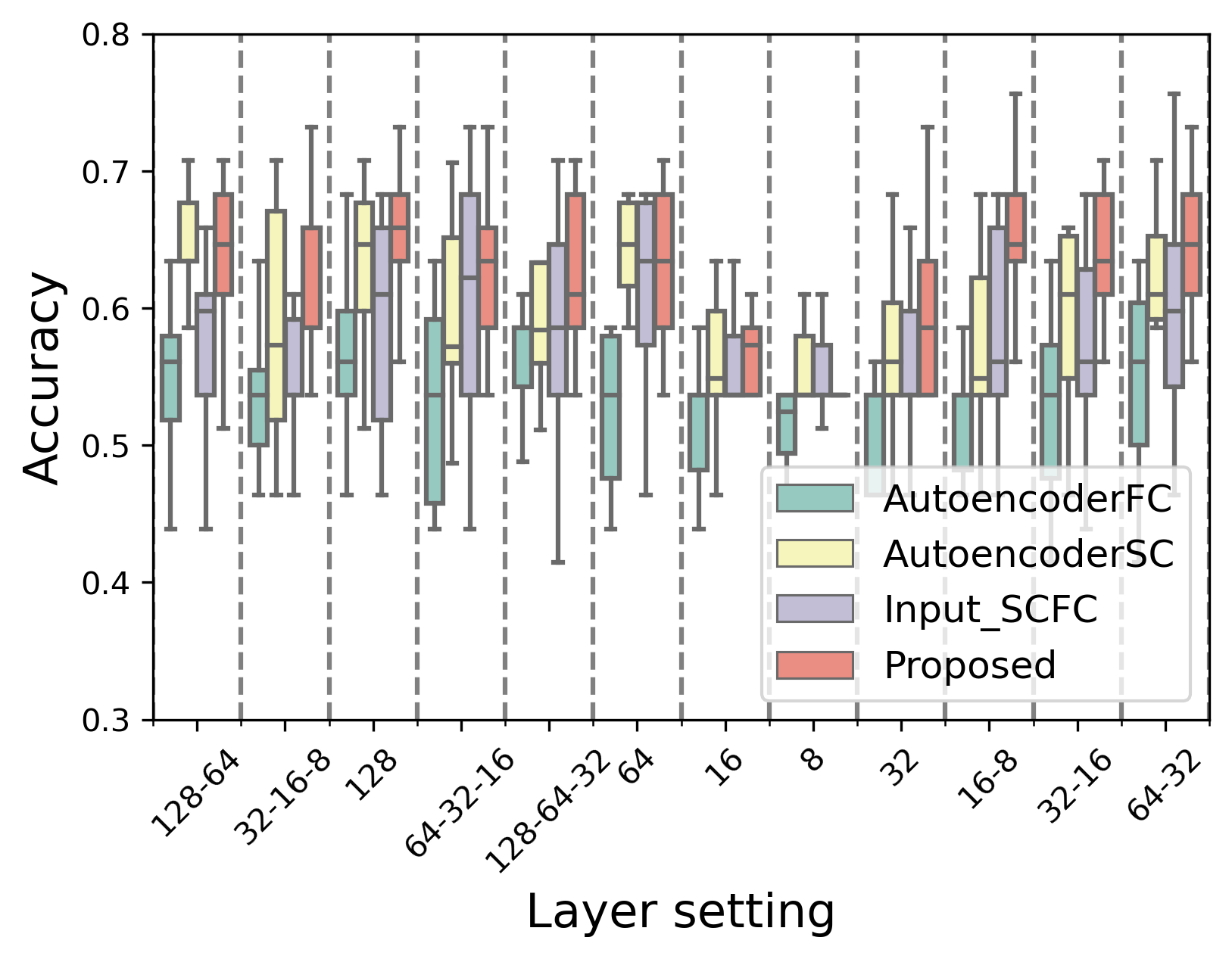}%
\label{fig:fcscacc}}
\hfil
{\includegraphics[height = 60mm, width=0.9\columnwidth]{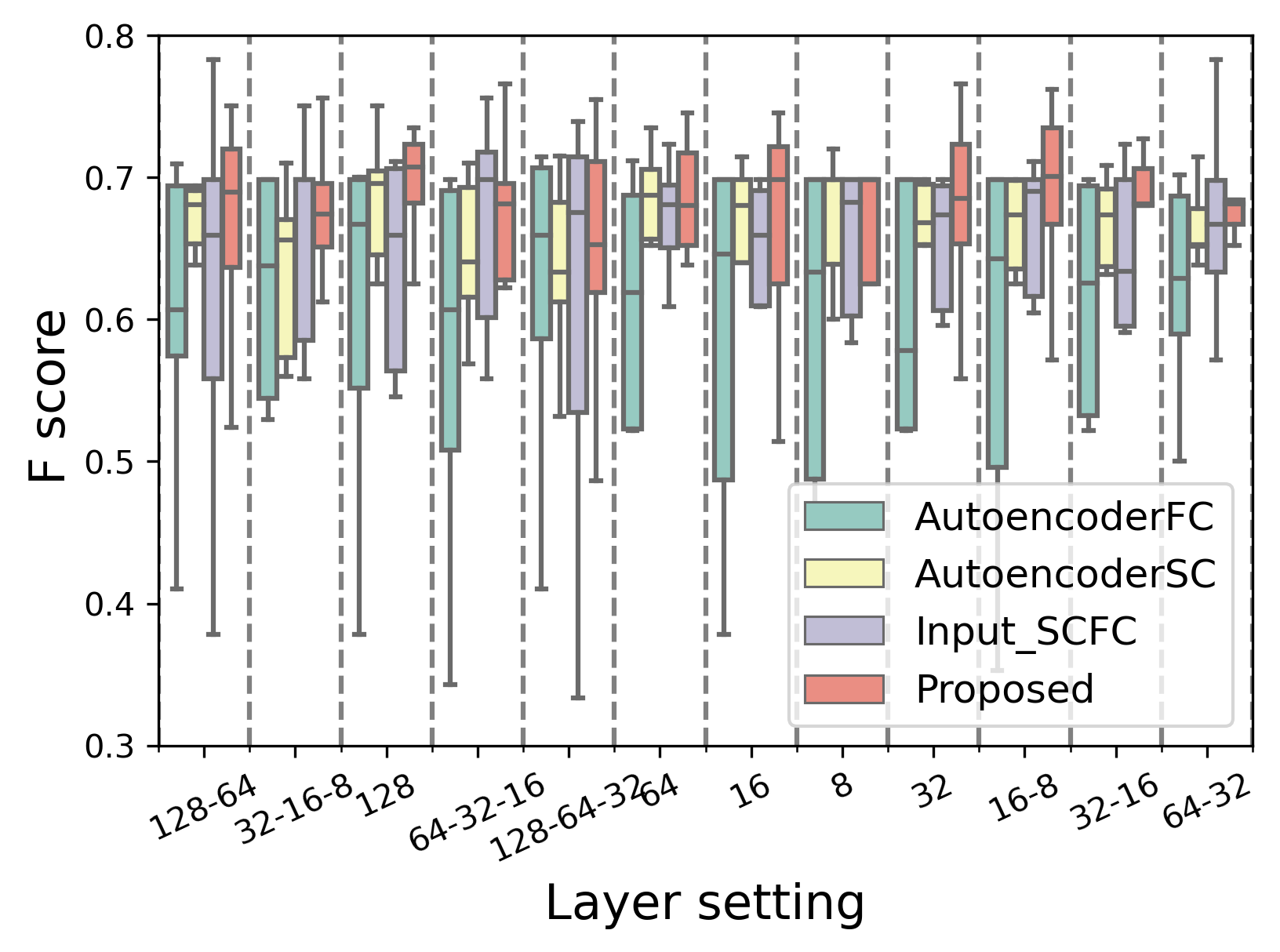}%
\label{fig:fcscfs}}
\caption{Classification performance comparison between proposed model and baselines. The proposed encoder-decoder model (shown in the rightmost red box) outperforms autoencoders that learn representations to reconstruct either FC or SC. It is also consistently superior to a baseline method where both SC and FC are fed as a combined input to a GCN-based feed-forward classifier.}
\label{fig:fscscompare}
\end{figure}

\noindent\textbf{Combined FC-SC input.} The proposed model learns the representation of the SC-FC relationship to distinguish heavy drinkers from non-drinkers. One valid question is whether using both SC and FC as the combined graph input to a feed-forward GCN-based classifier could also achieve competitive or similar performance. If so, there is no gain in using advanced GRL methods and considering the FC graph reconstruction pipeline as a regression component in the model. Moreover, this would imply that the SC to FC mapping we model is not as informative to distinguish among classes. In this comparative study we examine the aforementioned question. To this end, we combined both SC and FC networks of each subject into an $2N\times 2N$ adjacency matrix, where the upper left block corresponds to SC and lower right block contains the FC. The other entries are set to zero. The combined SC plus FC adjacency matrix is then used in the same GCN architectures as in Table~\ref{table_gcn} for a fair comparison. Figure~\ref{fig:fscscompare} depicts the classification performance comparison between all $12$ different GCN architectures ($L=1,2,3$ and different amounts of filters per layer), with node embedding concatenation and global average pooling. One finds that the proposed method outperforms this baseline in almost all of the architectural settings, and importantly this always holds for those models that yield the highest classification rates. The general trend remains the same without embedding concatenation (i.e., $\bbX_C=\bbX^{(L)}$), or, when using other global pooling schemes (max/sum).

\noindent\textbf{FC and SC autoencoders.} We also compared the proposed encoder-decoder system against two graph autoencoder baselines that individually process SC and FC, respectively.  Figure \ref{fig:fscscompare} shows that the proposed approach exhibits superior performance relative to the autoencoder baselines, both in terms of classification accuracy and F score. This offers further evidence supporting our hypothesis which states that, beyond the information encoded in the graph topology of each individual connectome (SC or FC), the relationship between them appears to provide additional insights and more discriminative learnt representations to improve subject classification (while capturing intrinsic properties of the brain structure-function coupling as a byproduct).

\begin{figure}[!t]
\centering
\subfloat[FC feature vector]{\includegraphics[height = 40mm, width=0.5\columnwidth]{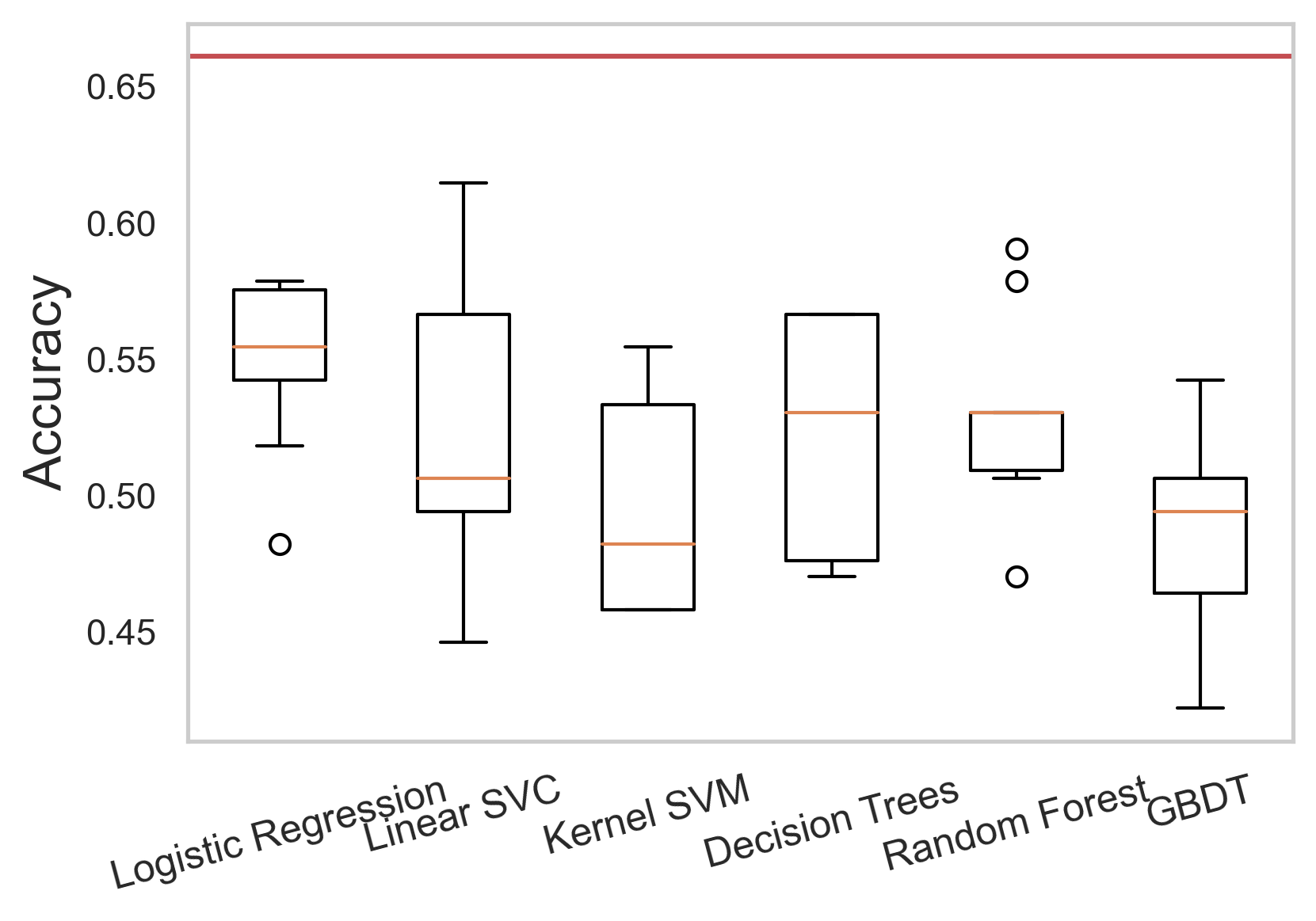}%
\includegraphics[height = 40mm, width=0.5\columnwidth]{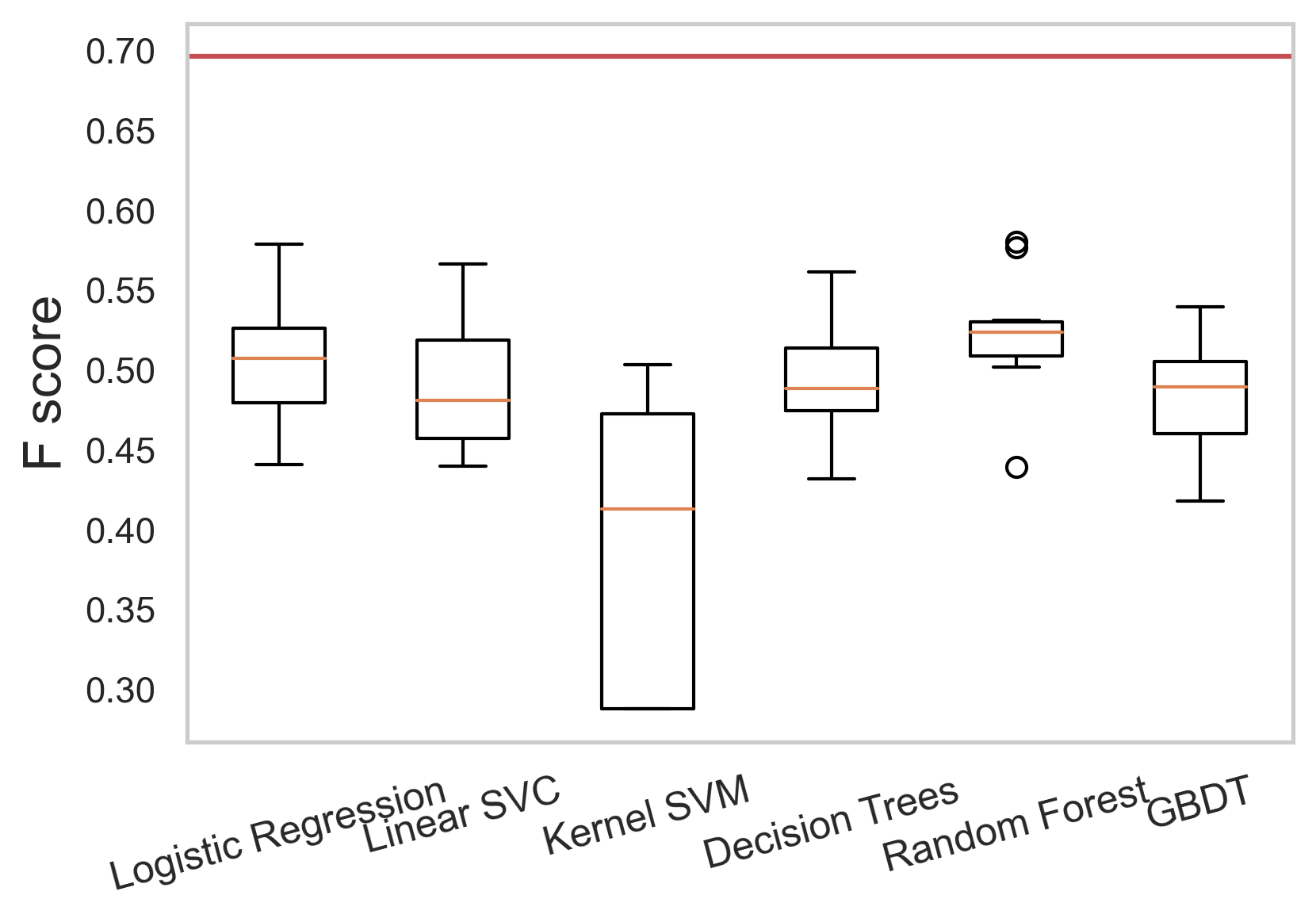}%
\label{fcml}}
\hfil
\subfloat[SC feature vector]{\includegraphics[height = 40mm, width=0.5\columnwidth]{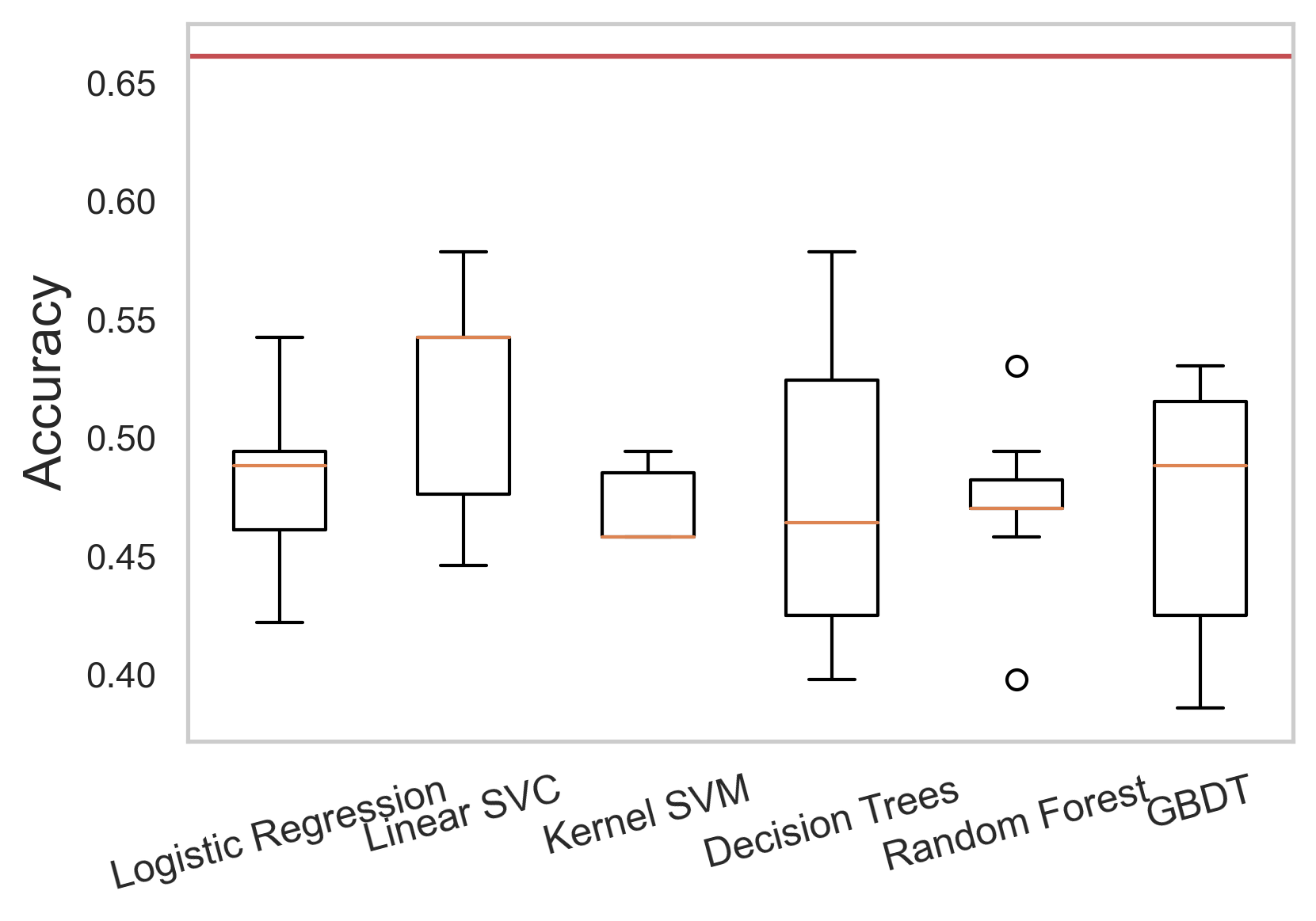}%
\includegraphics[height = 40mm, width=0.5\columnwidth]{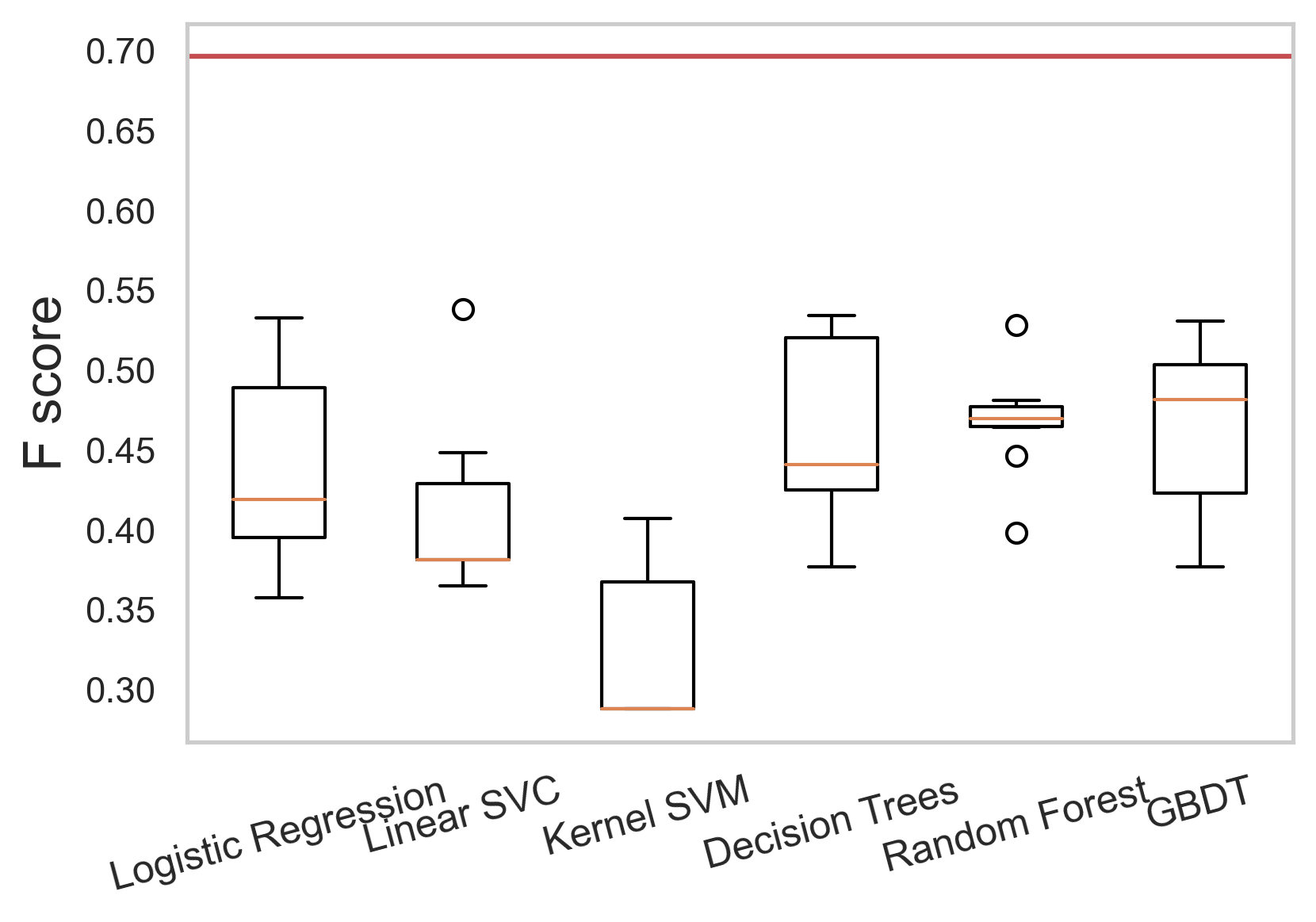}%
\label{scml}}
\hfil
\subfloat[FC + SC feature vector]{\includegraphics[height = 40mm, width=0.5\columnwidth]{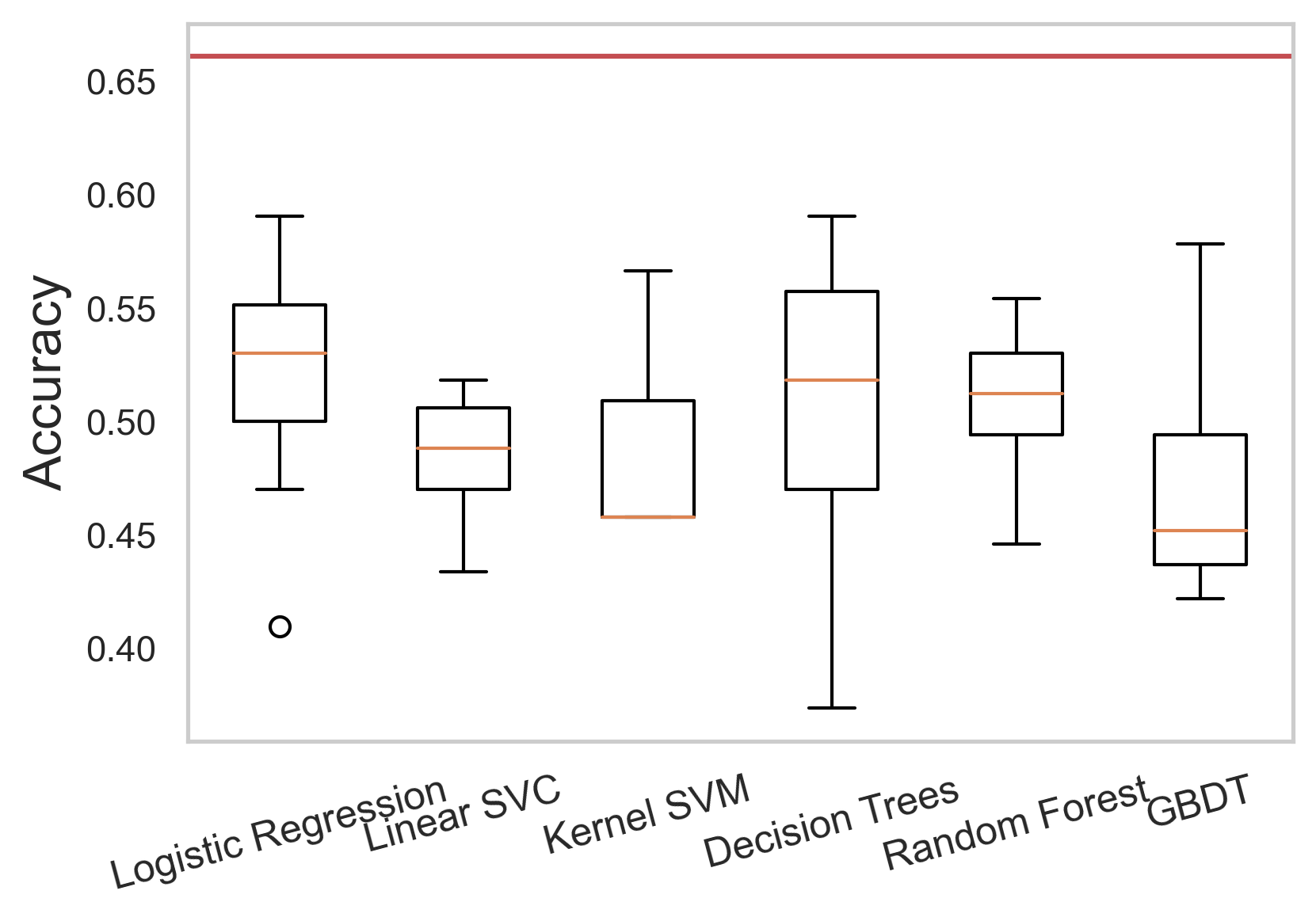}%
\includegraphics[height = 40mm, width=0.5\columnwidth]{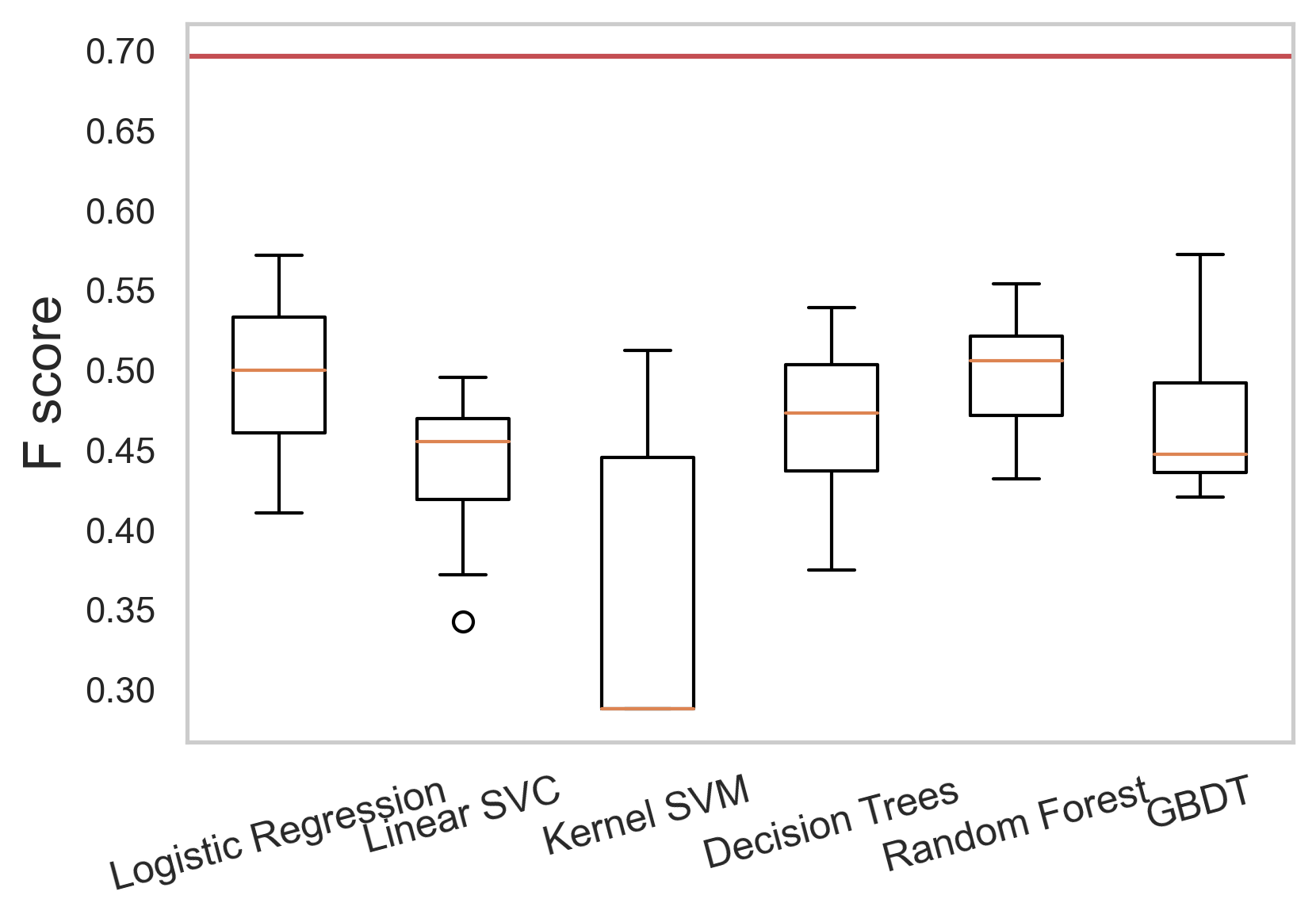}%
\label{fcscml}}
\caption{Classification performance of various ML models using handcrafted graph-theoretic feature vectors. The performance is not competitive with that of the proposed GRL encoder-decoder system marked by the red horizontal line in each subplot.}
\label{fig:ml}
\end{figure}

\noindent\textbf{ML with handcrafted graph-theoretic features.} Starting from both FC and SC connectomes, we compute network-centric measures that summarize different global graph properties such as modular structure, paths lengths, connectivity, and/or graph components as well as subgraph structures~\cite{rubinov2010complex,bullmore2009complex,kolaczyk2009book}. Specifically, we evaluate average path length, global efficiency, clustering coefficient, graph radius/diameter, transitivity, and graph density for the SC and FC networks of each subject~\cite{rubinov2010complex}. These summary statistics are collected in a feature vector representing the connectivity patterns of the subject's FC, SC, or both. Using these handcrafted feature vectors and the given labels, we train several classifiers for non-graph inputs. These include logistic regression, linear and kernel SVMs, decision trees, random forests and gradient-boosted decision trees (GBDTs). 
Results are presented in Figure \ref{fig:ml}. The conclusion is that using handcrafted features relative to learnt representations accounting for symmetries and invariances in graph data, leads to inferior classification performance. Needless to say, this principle has been verified in a multitude other domains and is at the heart of the deep learning revolution we have witnessed over the last dozen years.

\begin{figure}[!t]
\centering
\subfloat[]{\includegraphics[height = 40mm, width = 40mm]{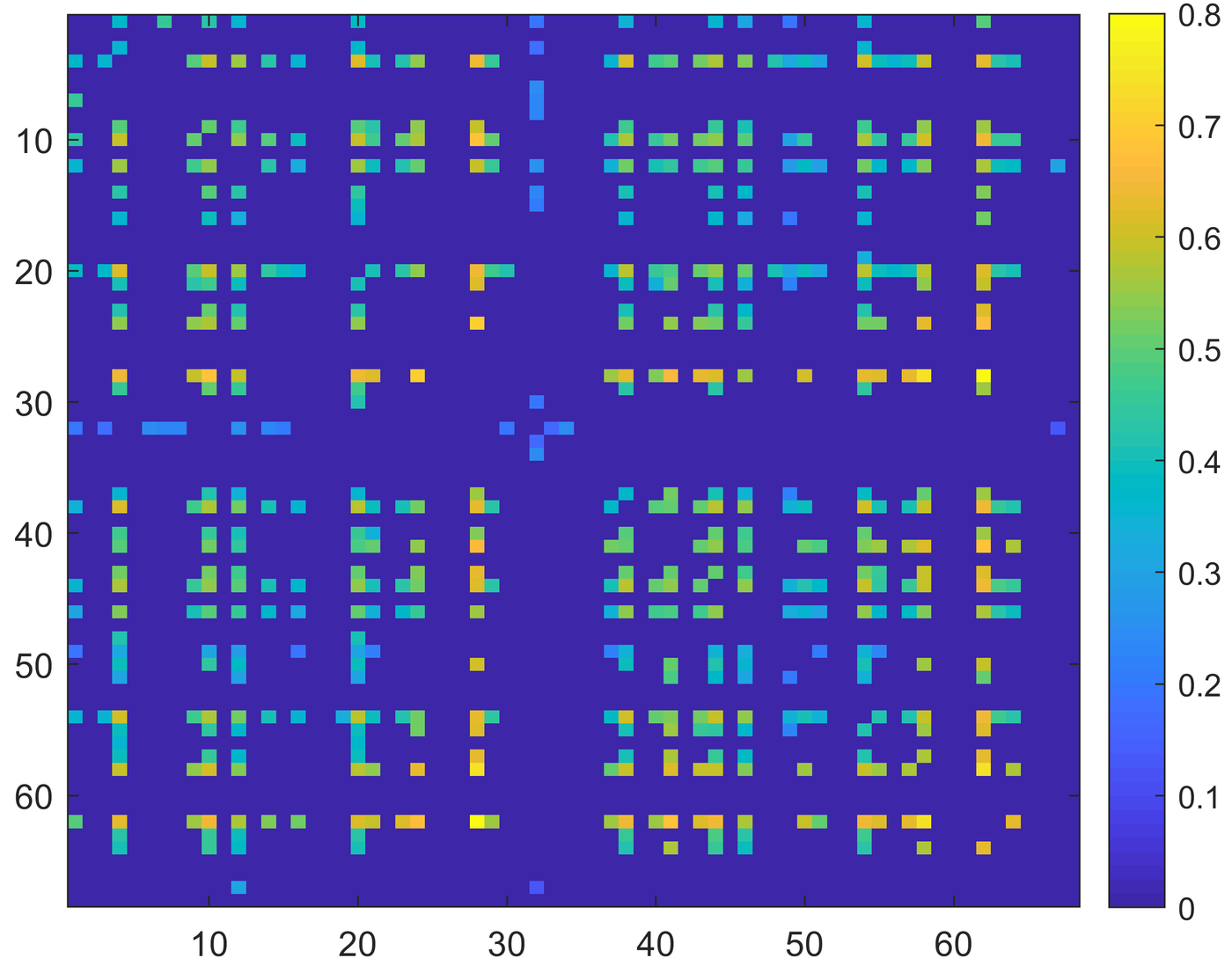}%
\label{fig_1}}
\hfil
\subfloat[]{\includegraphics[height = 40mm, width = 40mm]{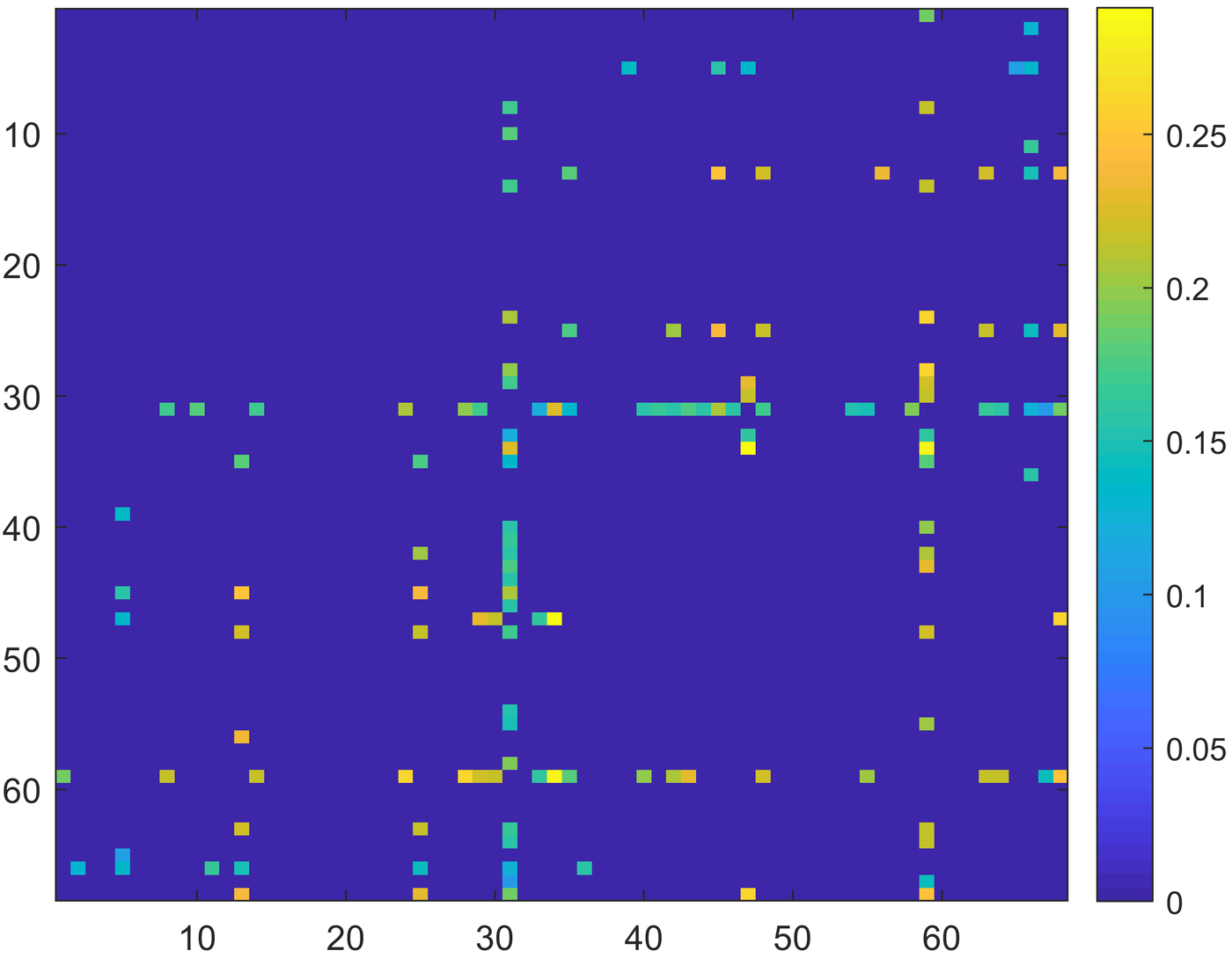}%
\label{fig_2}}
\hfil
\subfloat[]{\includegraphics[height = 40mm, width = 40mm]{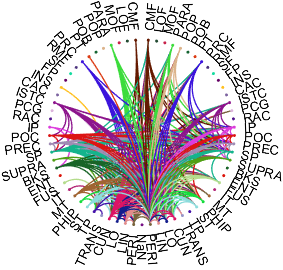}%
\label{fig_new1}}
\hfil
\subfloat[]{\includegraphics[height = 40mm, width = 40mm]{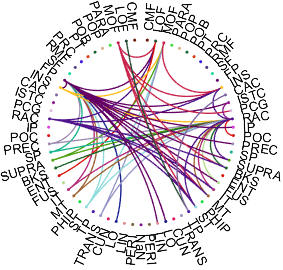}%
\label{fig_new2}}
\caption{Subgraphs with edges that were found to be significantly (a)(c) weaker, and (b)(d) stronger among the group of heavy drinkers ($p<0.05$). For visualization purposes, in the bottom plots edges are color-coded to match the color assigned to the source node. The difference in sparsity (edge densities) between the subgraphs in the (a)(c) and (b)(d) appears to suggest a marked weakening of functional links in heavy-drinkers, with some limited level of compensation via strengthened connections.}
\label{fig:fig_dissect}
\end{figure}

\subsection{Analysis on reconstructed FC networks}

By virtue of the combined loss function \eqref{E:loss_func} in the model, learnt nodal representations serve as discriminative features to distinguish among the two drinking-related classes. These representations are also used to reconstruct the FC networks $\hbSigma$, so we expect to see some of this embedded discriminability in the reconstructed graphs themselves. To examine and interpret these patterns, 
we test for significant between-group (i.e., heavy drinkers and non-drinkers) differences in the topology of the graphs $\hbSigma$. In more detail, for each potential functional connection $(i,j)\in\ccalE$, i.e., each element $\hat{\Sigma}_{ij}$ in the, say, upper-triangular half of $\hbSigma$, we perform a $t$-test with False Discovery Rate (FDR) correction for multiple testing. We report statistically-significant differences between groups at the $5\%$ level $(p < 0.05)$, and collections of significant links forming connected subgraphs are examined next. Results are shown in Figure~\ref{fig:fig_dissect}, where the top row depicts the adjacency matrix of the subgraphs and the lower row offers a visualization of the networks themselves.

Figures~\ref{fig_1} and~\ref{fig_new1} depict a subgraph of functional connections that are significantly weaker for the group of heavy drinkers.  For reference, edge weights in Figure~\ref{fig_1} correspond to the average functional connection strengths in the non-drinker group. Figure~\ref{fig_new1} shows an important connectivity decrease for brain RoIs that can be traced to the parietal, cingulate, temporal, occipital, and frontal cortices. The spatial breadth of the observed FC weakening suggests that alcohol may not only change localized regional brain activity, but also affect global patterns of brain functional organization; see~\cite{volkow2008moderate,oscar2003alcoholism} for related findings. 

Interestingly, the observed effect of drinking does not only reflect FC weakening and sparsification. A different subgraph shown in Figures~\ref{fig_2} and~\ref{fig_new2}, illustrates functional edges that were found to be significantly stronger in the group of heavy-drinkers. Nodes in said subgraph span various cortical RoIs and can be attributed to a neural compensation or equalization effect, which has been investigated extensively in the field of cognitive analysis~\cite{barulli2013efficiency}. Prior findings on neural compensation suggest it correlates with a combination of brain structure and function-related effects,  hence it is more likely to be revealed in models that jointly account for SC and FC as we do here. All in all, our results are aligned with the literature~\cite{parks2003brain} that reports the brain tends to establish additional connections to compensate for neuro-functional damage caused by alcohol. A deeper neuroscientific interpretation in the broader context of robustness and resilience of human brain activity is well beyond the scope of this methodological paper. Nonetheless we believe these analyses can be valuable from the perspective of explainability, further motivating the application of GRL advances for neuroimaging data research.


\section{Concluding Summary and Outlook}\label{sec:conclusion}
We developed a GCN-based encoder-decoder system to learn low-dimensional embeddings that carry information about the relationship between brain structural and functional connectomes. Although there have been ample SC- and FC-based studies to investigate the difference between brain connectomes of patients and healthy controls, using GRL to model the structure-function coupling for subject (i.e., graph) classification has been unexplored so far. By putting forth a flexible multi-objective framework the learnt representations can be used in various types of downstream tasks, thus markedly broadening the range of network neuroscience applications. In the proposed model, two learning objectives are considered. Given SC networks as inputs, the first objective is to reconstruct the FC of the same subject as a regression problem. The second objective is to generate graph-level embeddings that capture individual subject's variability and are discriminative among the classes of heavy drinkers and non-drinkers. This enables supervised subject-level classification. 

We conducted a rather broad exploration of the architectural design space. Specifically, we investigated a wide range of GCN models and also tested different global pooling mechanisms. For deep GCNs, nodal embeddings that concatenate the output of earlier layers were shown to improve performance in both graph reconstruction and subject classification. Through experimental comparisons with a broad lineup of baseline methods, we showed that: (i) end-to-end multi-objective training attains better classification performance than a two-step method where graph reconstruction and classification are trained separately; and (ii) the proposed GRL model could extract succinct, yet informative latent representations of the SC-FC mapping. Relative to baseline models that learn from SC, FC (using autoencoders), or both networks fed as inputs to a feed-forward GCN, the obtained SC-FC representations offer additional information, leading to higher classification accuracy and F score. Moreover, (iii) the proposed method outperformed ML models trained on hand-crafted features collecting various graph-theoretic summary statistics computed from the FC and SC networks. This highlights the superior discriminative power of representations learnt using  geometric deep learning, which favorably exploit symmetries and invariances in  graph data. Finally, from the reconstructed FC for all subjects, statistically-significant differences in edges between drinkers and non-drinkers were revealed. Our findings suggest a marked weakening of functional links in heavy-drinkers, with some limited level of neural compensation in the form of sparse strengthened connections. 

This being the first work to build a GRL model of SC-FC brain networks, several extensions are possible. For example, we used one-hot encoding as the initial nodal attribute for each node. An open challenge concerns situations where nodal attributes can be collected from meaningful subject-related information, such as the physical volume of RoIs. In the convolutional layers of the encoder, one could also adopt higher-order graph filters to further gain in expressive power. Node- or edge-varying graph filters are other valuable alternatives in this direction, possibly departing from convolutional models. Also, attention mechanisms can also be applied for the global pooling. Our focus here was on a simple, yet powerful architecture to demonstrate conceptual advantages of the model -- not to push e.g., subject classification accuracy to the limit. In a broader context, a similar encoder-decoder system (modulo application-specific modifications) could be applied to other domains where there exists an intrinsic relationship between two networks. Examples include snapshots of traffic networks, dynamic brain connectomes at consecutive time stamps, or sensor-collected human co-location graphs along with Facebook ties for friendship recommendation via link prediction.

\ifCLASSOPTIONcaptionsoff
  \newpage
\fi



%

\bibliographystyle{IEEEtran}
\bibliography{mycitations}

\begin{thebibliography}{10}
\providecommand{\url}[1]{#1}
\csname url@samestyle\endcsname
\providecommand{\newblock}{\relax}
\providecommand{\bibinfo}[2]{#2}
\providecommand{\BIBentrySTDinterwordspacing}{\spaceskip=0pt\relax}
\providecommand{\BIBentryALTinterwordstretchfactor}{4}
\providecommand{\BIBentryALTinterwordspacing}{\spaceskip=\fontdimen2\font plus
\BIBentryALTinterwordstretchfactor\fontdimen3\font minus
  \fontdimen4\font\relax}
\providecommand{\BIBforeignlanguage}[2]{{%
\expandafter\ifx\csname l@#1\endcsname\relax
\typeout{** WARNING: IEEEtran.bst: No hyphenation pattern has been}%
\typeout{** loaded for the language `#1'. Using the pattern for}%
\typeout{** the default language instead.}%
\else
\language=\csname l@#1\endcsname
\fi
#2}}
\providecommand{\BIBdecl}{\relax}
\BIBdecl

\bibitem{li2020supervised}
Y.~Li, R.~Shafipour, G.~Mateos, and Z.~Zhang, ``Supervised graph representation
  learning for modeling the relationship between structural and functional
  brain connectivity,'' \emph{IEEE Intl. Conf. Acoust., Speech and Signal
  Process. (ICASSP)}, pp. 9065--9069, 2020.

\bibitem{van2010exploring}
M.~Van Den~Heuvel and H.~Pol, ``Exploring the brain network: a review on
  resting-state f{MRI} functional connectivity,'' \emph{Eur.
  Neuropsychopharmacol.}, vol.~20, no.~8, pp. 519--534, 2010.

\bibitem{fornito2016fundamentals}
A.~Fornito, A.~Zalesky, and E.~Bullmore, \emph{Fundamentals of Brain Network
  Analysis}.\hskip 1em plus 0.5em minus 0.4em\relax Academic Press, 2016.

\bibitem{power2011functional}
J.~D. Power, A.~L. Cohen, S.~M. Nelson, G.~S. Wig, K.~A. Barnes, J.~A. Church,
  A.~C. Vogel, T.~O. Laumann, F.~M. Miezin, B.~L. Schlaggar \emph{et~al.},
  ``Functional network organization of the human brain,'' \emph{Neuron},
  vol.~72, no.~4, pp. 665--678, 2011.

\bibitem{bullmore2009complex}
E.~Bullmore and O.~Sporns, ``Complex brain networks: {G}raph theoretical
  analysis of structural and functional systems,'' \emph{Nat. Rev. Neurosci.},
  vol.~10, no.~3, p. 186, 2009.

\bibitem{sporns2010networks}
O.~Sporns, \emph{Networks of the Brain}.\hskip 1em plus 0.5em minus 0.4em\relax
  MIT Press, 2010.

\bibitem{abdelnour2014network}
F.~Abdelnour, H.~U. Voss, and A.~Raj, ``Network diffusion accurately models the
  relationship between structural and functional brain connectivity networks,''
  \emph{Neuroimage}, vol.~90, pp. 335--347, 2014.

\bibitem{li2019identifying}
Y.~Li and G.~Mateos, ``Identifying structural brain networks from functional
  connectivity: A network deconvolution approach,'' \emph{IEEE Intl. Conf.
  Acoust., Speech and Signal Process. (ICASSP)}, pp. 1135--1139, 2019.

\bibitem{honey2009predicting}
C.~Honey, O.~Sporns, L.~Cammoun, X.~Gigandet, J.-P. Thiran, R.~Meuli, and
  P.~Hagmann, ``Predicting human resting-state functional connectivity from
  structural connectivity,'' \emph{Proc. Natl. Acad. Sci. U.S.A.}, vol. 106,
  no.~6, pp. 2035--2040, 2009.

\bibitem{goni2014resting}
J.~Go{\~n}i, M.~Van Den~Heuvel, A.~Avena-Koenigsberger, N.~V. de~Mendizabal,
  R.~F. Betzel, A.~Griffa, P.~Hagmann, B.~Corominas-Murtra, J.-P. Thiran, and
  O.~Sporns, ``Resting-brain functional connectivity predicted by analytic
  measures of network communication,'' \emph{Proc. Natl. Acad. Sci. U.S.A.},
  vol. 111, no.~2, pp. 833--838, 2014.

\bibitem{stam2016relation}
C.~Stam, E.~Van~Straaten, E.~Van~Dellen, P.~Tewarie, G.~Gong, A.~Hillebrand,
  J.~Meier, and P.~Van~Mieghem, ``The relation between structural and
  functional connectivity patterns in complex brain networks,'' \emph{Intl. J.
  Psychophysiol.}, vol. 103, pp. 149--160, 2016.

\bibitem{honey2010can}
C.~J. Honey, J.-P. Thivierge, and O.~Sporns, ``Can structure predict function
  in the human brain?'' \emph{Neuroimage}, vol.~52, no.~3, pp. 766--776, 2010.

\bibitem{GlobalSIP19_YangLiEncoderDecoder}
Y.~Li, R.~Shafipour, G.~Mateos, and Z.~Zhang, ``Mapping brain structural
  connectivities to functional networks via graph encoder-decoder with
  interpretable latent embeddings,'' \emph{IEEE Global Conf. Signal and Info.
  Process. (GlobalSIP)}, pp. 1--5, 2019.

\bibitem{sarwar2021structure}
T.~Sarwar, Y.~Tian, B.~T. Yeo, K.~Ramamohanarao, and A.~Zalesky,
  ``Structure-function coupling in the human connectome: A machine learning
  approach,'' \emph{NeuroImage}, vol. 226, p. 117609, 2021.

\bibitem{kipf2016semi}
T.~N. Kipf and M.~Welling, ``Semi-supervised classification with graph
  convolutional networks,'' in \emph{Intl. Conf. on Learning Representations
  (ICLR)}, 2017.

\bibitem{duvenaud2015convolutional}
D.~Duvenaud, D.~Maclaurin, J.~Aguilera-Iparraguirre, R.~G{\'o}mez-Bombarelli,
  T.~Hirzel, A.~Aspuru-Guzik, and R.~P. Adams, ``Convolutional networks on
  graphs for learning molecular fingerprints,'' in \emph{Adv. Neural. Inf.
  Process. Syst. (NeurIPS)}, 2015, pp. 2224--2232.

\bibitem{gama2018convolutional}
F.~Gama, A.~G. Marques, G.~Leus, and A.~Ribeiro, ``Convolutional neural network
  architectures for signals supported on graphs,'' \emph{IEEE Trans. Signal
  Process.}, vol.~67, no.~4, pp. 1034--1049, 2018.

\bibitem{ying2018graph}
R.~Ying, R.~He, K.~Chen, P.~Eksombatchai, W.~L. Hamilton, and J.~Leskovec,
  ``Graph convolutional neural networks for web-scale recommender systems,'' in
  \emph{ACM SIGKDD Intl. Conf. on Knowledge Discovery and Data Mining (KDD)},
  2018, pp. 974--983.

\bibitem{hamilton2017inductive}
W.~L. Hamilton, R.~Ying, and J.~Leskovec, ``Inductive representation learning
  on large graphs,'' in \emph{Adv. Neural. Inf. Process. Syst. (NeurIPS)},
  2017, pp. 1025--1035.

\bibitem{yue2020graph}
X.~Yue, Z.~Wang, J.~Huang, S.~Parthasarathy, S.~Moosavinasab, Y.~Huang, S.~Lin,
  W.~Zhang, P.~Zhang, and H.~Sun, ``Graph embedding on biomedical networks:
  methods, applications and evaluations.'' \emph{Bioinformatics}, vol.~36,
  no.~4, pp. 1241--1251, 2020.

\bibitem{ktena2018metric}
S.~I. Ktena, S.~Parisot, E.~Ferrante, M.~Rajchl, M.~Lee, B.~Glocker, and
  D.~Rueckert, ``Metric learning with spectral graph convolutions on brain
  connectivity networks,'' \emph{NeuroImage}, vol. 169, pp. 431--442, 2018.

\bibitem{kipf2016variational}
T.~N. Kipf and M.~Welling, ``Variational graph auto-encoders,'' \emph{NeurIPS
  Workshop on Bayesian Deep Learning}, 2016.

\bibitem{chami2020machine}
I.~Chami, S.~Abu-El-Haija, B.~Perozzi, C.~R{\'e}, and K.~Murphy, ``Machine
  learning on graphs: A model and comprehensive taxonomy,'' \emph{arXiv
  preprint arXiv:2005.03675}, 2020.

\bibitem{cheung2019pooling}
M.~Cheung, J.~Shi, L.~Jiang, O.~Wright, and J.~M. Moura, ``Pooling in graph
  convolutional neural networks,'' in \emph{Asilomar Conf. on Signals, Systems,
  and Computers}, 2019, pp. 462--466.

\bibitem{HCP}
``Connectome {C}oordination {F}acility,''
  \url{https://www.humanconnectome.org/}, accessed: 2021-10-12.

\bibitem{gama2020spmag}
F.~Gama, E.~Isufi, G.~Leus, and A.~Ribeiro, ``Graphs, convolutions, and neural
  networks: From graph filters to graph neural networks,'' \emph{IEEE Signal
  Process. Mag.}, vol.~37, no.~6, pp. 128--138, 2020.

\bibitem{bronstein2017geometric}
M.~M. Bronstein, J.~Bruna, Y.~LeCun, A.~Szlam, and P.~Vandergheynst,
  ``Geometric deep learning: going beyond {E}uclidean data,'' \emph{IEEE Signal
  Process. Mag.}, vol.~34, no.~4, pp. 18--42, 2017.

\bibitem{errica2020fair}
F.~Errica, M.~Podda, D.~Bacciu, and A.~Micheli, ``A fair comparison of graph
  neural networks for graph classification,'' in \emph{Intl. Conf. on Learning
  Representations (ICLR)}, 2020.

\bibitem{wu2021comprehensive}
Z.~Wu, S.~Pan, F.~Chen, G.~Long, C.~Zhang, and S.~Y. Philip, ``A comprehensive
  survey on graph neural networks,'' \emph{IEEE Trans. Neural Netw. Learn.
  Syst}, vol.~32, no.~1, pp. 4--24, 2021.

\bibitem{bruna2013spectral}
J.~Bruna, W.~Zaremba, A.~Szlam, and Y.~LeCun, ``Spectral networks and locally
  connected networks on graphs,'' in \emph{Intl. Conf. on Learning
  Representations (ICLR)}, 2014.

\bibitem{xu2018powerful}
K.~Xu, W.~Hu, J.~Leskovec, and S.~Jegelka, ``How powerful are graph neural
  networks?'' in \emph{Intl. Conf. on Learning Representations (ICLR)}, 2019.

\bibitem{hamilton2020book}
W.~L. Hamilton, ``Graph representation learning,'' \emph{Synth. Lect. Artif.
  Intell. Mach. Learn.}, vol.~14, no.~3, pp. 1--159, 2020.

\bibitem{hamilton2017representation}
W.~L. Hamilton, R.~Ying, and J.~Leskovec, ``Representation learning on graphs:
  Methods and applications,'' \emph{IEEE Data Engineering Bulletin}, 2017.

\bibitem{gutierrez2019unsupervised}
L.~Guti{\'e}rrez-G{\'o}mez and J.-C. Delvenne, ``Unsupervised network embedding
  for graph visualization, clustering and classification,'' \emph{arXiv
  preprint arXiv:1903.05980}, 2019.

\bibitem{zhang2018linkpred}
M.~Zhang and Y.~Chen, ``Link prediction based on graph neural networks,'' in
  \emph{Adv. Neural. Inf. Process. Syst. (NeurIPS)}, vol.~31, 2018.

\bibitem{tsitsulin2020graph}
A.~Tsitsulin, J.~Palowitch, B.~Perozzi, and E.~M{\"u}ller, ``Graph clustering
  with graph neural networks,'' \emph{arXiv preprint arXiv:2006.16904}, 2020.

\bibitem{tian2014learning}
F.~Tian, B.~Gao, Q.~Cui, E.~Chen, and T.-Y. Liu, ``Learning deep
  representations for graph clustering,'' in \emph{Proc. AAAI Conf. on Artif.
  Intell.}, vol.~28, no.~1, 2014.

\bibitem{narayanan2017graph2vec}
A.~Narayanan, M.~Chandramohan, R.~Venkatesan, L.~Chen, Y.~Liu, and S.~Jaiswal,
  ``graph2vec: Learning distributed representations of graphs,'' \emph{arXiv
  preprint arXiv:1707.05005}, 2017.

\bibitem{wang2021generalizable}
P.~Y. Wang, S.~Sapra, V.~K. George, and G.~A. Silva, ``Generalizable machine
  learning in neuroscience using graph neural networks,'' \emph{Front. Artif.
  Intell.}, vol.~4, p.~4, 2021.

\bibitem{rosenthal2018mapping}
G.~Rosenthal, F.~V{\'a}{\v{s}}a, A.~Griffa, P.~Hagmann, E.~Amico, J.~Go{\~n}i,
  G.~Avidan, and O.~Sporns, ``Mapping higher-order relations between brain
  structure and function with embedded vector representations of connectomes,''
  \emph{Nat. Commun.}, vol.~9, no.~1, p. 2178, 2018.

\bibitem{kim2020understanding}
B.-H. Kim and J.~C. Ye, ``Understanding graph isomorphism network for brain
  {MR} functional connectivity analysis,'' \emph{arXiv preprint
  arXiv:2001.03690}, 2020.

\bibitem{zhang2018multi}
X.~Zhang, L.~He, K.~Chen, Y.~Luo, J.~Zhou, and F.~Wang, ``Multi-view graph
  convolutional network and its applications on neuroimage analysis for
  parkinson’s disease,'' in \emph{AMIA Annu. Symp. Proc.}, vol. 2018, 2018,
  p. 1147.

\bibitem{parisot2018disease}
S.~Parisot, S.~I. Ktena, E.~Ferrante, M.~Lee, R.~Guerrero, B.~Glocker, and
  D.~Rueckert, ``Disease prediction using graph convolutional networks:
  application to autism spectrum disorder and {A}lzheimer’s disease,''
  \emph{Med. Image Anal.}, vol.~48, pp. 117--130, 2018.

\bibitem{gilmer2017neural}
J.~Gilmer, S.~S. Schoenholz, P.~F. Riley, O.~Vinyals, and G.~E. Dahl, ``Neural
  message passing for quantum chemistry,'' in \emph{Intl. Conf. on Machine
  Learning (ICML)}, 2017, pp. 1263--1272.

\bibitem{garg2020generalization}
V.~Garg, S.~Jegelka, and T.~Jaakkola, ``Generalization and representational
  limits of graph neural networks,'' in \emph{Intl. Conf. on Machine Learning
  (ICML)}, 2020, pp. 3419--3430.

\bibitem{ortega2018graph}
A.~Ortega, P.~Frossard, J.~Kovačević, J.~M.~F. Moura, and P.~Vandergheynst,
  ``Graph signal processing: Overview, challenges, and applications,''
  \emph{Proc. IEEE}, vol. 106, no.~5, pp. 808--828, 2018.

\bibitem{huang2018graph}
W.~Huang, T.~A. Bolton, J.~D. Medaglia \emph{et~al.}, ``A graph signal
  processing perspective on functional brain imaging,'' \emph{Proc. IEEE}, vol.
  106, no.~5, 2018.

\bibitem{defferrard2016convolutional}
M.~Defferrard, X.~Bresson, and P.~Vandergheynst, ``Convolutional neural
  networks on graphs with fast localized spectral filtering,'' in \emph{Adv.
  Neural. Inf. Process. Syst. (NeurIPS)}, 2016, pp. 3844--3852.

\bibitem{you2020design}
J.~You, Z.~Ying, and J.~Leskovec, ``Design space for graph neural networks,''
  in \emph{Adv. Neural. Inf. Process. Syst. (NeurIPS)}, vol.~33, 2020.

\bibitem{van2013wu}
D.~C. Van~Essen, S.~M. Smith, D.~M. Barch, T.~E. Behrens, E.~Yacoub,
  K.~Ugurbil, W.-M.~H. Consortium \emph{et~al.}, ``The {W}u-{M}inn human
  connectome project: an overview,'' \emph{Neuroimage}, vol.~80, pp. 62--79,
  2013.

\bibitem{glasser2016human}
M.~F. Glasser, S.~M. Smith, D.~S. Marcus, J.~L. Andersson, E.~J. Auerbach,
  T.~E. Behrens, T.~S. Coalson, M.~P. Harms, M.~Jenkinson, S.~Moeller
  \emph{et~al.}, ``The human connectome project's neuroimaging approach,''
  \emph{Nat. Neurosci.}, vol.~19, no.~9, pp. 1175--1187, 2016.

\bibitem{desikan2006automated}
R.~S. Desikan, F.~S{\'e}gonne, B.~Fischl, B.~T. Quinn, B.~C. Dickerson,
  D.~Blacker, R.~L. Buckner, A.~M. Dale, R.~P. Maguire, B.~T. Hyman
  \emph{et~al.}, ``An automated labeling system for subdividing the human
  cerebral cortex on {MRI} scans into gyral based regions of interest,''
  \emph{Neuroimage}, vol.~31, no.~3, pp. 968--980, 2006.

\bibitem{zhang2019tensor}
Z.~Zhang, G.~I. Allen, H.~Zhu, and D.~Dunson, ``Tensor network factorizations:
  Relationships between brain structural connectomes and traits,''
  \emph{Neuroimage}, vol. 197, pp. 330--343, 2019.

\bibitem{zhang2018mapping}
Z.~Zhang, M.~Descoteaux, J.~Zhang, G.~Girard, M.~Chamberland, D.~Dunson,
  A.~Srivastava, and H.~Zhu, ``Mapping population-based structural
  connectomes,'' \emph{NeuroImage}, vol. 172, pp. 130--145, 2018.

\bibitem{richiardi2013machine}
J.~Richiardi, S.~Achard, H.~Bunke, and D.~Van De~Ville, ``Machine learning with
  brain graphs: predictive modeling approaches for functional imaging in
  systems neuroscience,'' \emph{IEEE Signal Process. Mag.}, vol.~30, no.~3, pp.
  58--70, 2013.

\bibitem{power2010development}
J.~D. Power, D.~A. Fair, B.~L. Schlaggar, and S.~E. Petersen, ``The development
  of human functional brain networks,'' \emph{Neuron}, vol.~67, no.~5, pp.
  735--748, 2010.

\bibitem{rubinov2010complex}
M.~Rubinov and O.~Sporns, ``Complex network measures of brain connectivity:
  uses and interpretations,'' \emph{Neuroimage}, vol.~52, no.~3, pp.
  1059--1069, 2010.

\bibitem{bai2019unsupervised}
Y.~Bai, H.~Ding, Y.~Qiao, A.~Marinovic, K.~Gu, T.~Chen, Y.~Sun, and W.~Wang,
  ``Unsupervised inductive graph-level representation learning via graph-graph
  proximity,'' \emph{Intl. Joint Conf. Artif. Intell.}, pp. 1988--1994, 2019.

\bibitem{maas2013rectifier}
A.~L. Maas, A.~Y. Hannun, and A.~Y. Ng, ``Rectifier nonlinearities improve
  neural network acoustic models,'' in \emph{Intl. Conf. on Machine Learning
  (ICML)}, vol.~30, no.~1, 2013.

\bibitem{dai2016discriminative}
H.~Dai, B.~Dai, and L.~Song, ``Discriminative embeddings of latent variable
  models for structured data,'' in \emph{Intl. Conf. on Machine Learning
  (ICML)}, 2016, pp. 2702--2711.

\bibitem{abadi2016tensorflow}
M.~Abadi, P.~Barham, J.~Chen, Z.~Chen, A.~Davis, J.~Dean, M.~Devin,
  S.~Ghemawat, G.~Irving, M.~Isard \emph{et~al.}, ``Tensorflow: A system for
  large-scale machine learning,'' in \emph{{P}roc {USENIX} Symp. Oper. Syst.
  Des. Implement. (OSDI)}, 2016, pp. 265--283.

\bibitem{glorot2010understanding}
X.~Glorot and Y.~Bengio, ``Understanding the difficulty of training deep feed
  forward neural networks,'' in \emph{Intl. Conf. on Artif. Intell. Stat.},
  2010, pp. 249--256.

\bibitem{kingma2014adam}
D.~P. Kingma and J.~Ba, ``Adam: A method for stochastic optimization,'' in
  \emph{Intl. Conf. on Learning Representations (ICLR)}, 2015.

\bibitem{weston2012deep}
J.~Weston, F.~Ratle, H.~Mobahi, and R.~Collobert, ``Deep learning via
  semi-supervised embedding,'' in \emph{Neural networks: Tricks of the
  trade}.\hskip 1em plus 0.5em minus 0.4em\relax Springer, 2012, pp. 639--655.

\bibitem{kolaczyk2009book}
E.~D. Kolaczyk, \emph{Statistical Analysis of Network Data: Methods and
  Models}.\hskip 1em plus 0.5em minus 0.4em\relax New York, NY: Springer, 2009.

\bibitem{elements_of_statistics}
T.~Hastie, R.~Tibshirani, and J.~Friedman, \emph{The Elements of Statistical
  Learning}, 2nd~ed.\hskip 1em plus 0.5em minus 0.4em\relax Springer, 2009.

\bibitem{maaten2008visualizing}
L.~Maaten and G.~Hinton, ``Visualizing data using t-{SNE},'' \emph{J. Mach.
  Learn. Res.}, vol.~9, pp. 2579--2605, 2008.

\bibitem{volkow2008moderate}
N.~D. Volkow, Y.~Ma, W.~Zhu, J.~S. Fowler, J.~Li, M.~Rao, K.~Mueller,
  K.~Pradhan, C.~Wong, and G.-J. Wang, ``Moderate doses of alcohol disrupt the
  functional organization of the human brain,'' \emph{Psychiatry Res.
  Neuroimaging}, vol. 162, no.~3, pp. 205--213, 2008.

\bibitem{oscar2003alcoholism}
M.~Oscar-Berman and K.~Marinkovic, ``Alcoholism and the brain: {A}n overview,''
  \emph{Alcohol Res. Health}, vol.~27, no.~2, pp. 125--133, 2003.

\bibitem{barulli2013efficiency}
D.~Barulli and Y.~Stern, ``Efficiency, capacity, compensation, maintenance,
  plasticity: {E}merging concepts in cognitive reserve,'' \emph{Trends Cogn.
  Sci.}, vol.~17, no.~10, pp. 502--509, 2013.

\bibitem{parks2003brain}
M.~H. Parks, V.~L. Morgan, D.~R. Pickens, R.~R. Price, M.~S. Dietrich, M.~K.
  Nickel, and P.~R. Martin, ``Brain f{MRI} activation associated with
  self-paced finger tapping in chronic alcohol-dependent patients,''
  \emph{Alcohol.: Clin. Exp. Res.}, vol.~27, no.~4, pp. 704--711, 2003.

\end{thebibliography}



%








\end{document}